\documentclass[a4paper]{aa}

\usepackage{txfonts}
\usepackage{graphicx}
\usepackage{subfigure}
\usepackage{natbib}
\usepackage{hhline}
\usepackage{epsf}
\usepackage{lscape}
\usepackage{rotating}
\usepackage{tabularx}
\usepackage{lscape}
\usepackage{supertabular}
\usepackage{multirow}
\usepackage{wasysym}
\usepackage{hyperref}
\newcommand{\mic}{\,{\rm \mu m} }
\newcommand{\be}{\begin{equation}}
\newcommand{\ee}{\end{equation}}
\newcommand{\bea}{\begin{eqnarray}}
\newcommand{\eea}{\end{eqnarray}}
\newcommand{\ba}{\begin{array}}
\newcommand{\ea}{\end{array}}

\begin{document}   

\title{An analysis of star formation with Herschel in the Hi-GAL Survey.
II. The tips of the Galactic bar. }

\author{M. Veneziani\inst{1}, E. Schisano\inst{1,2}, D. Elia\inst{2}, A. Noriega-Crespo\inst{1,3}, S. Carey\inst{1}, A. Di Giorgio\inst{2}, Y. Fukui\inst{4}, B.M.T. Maiolo\inst{5,6}, Y. Maruccia\inst{5,6}, A. Mizuno\inst{7}, N. Mizuno\inst{8}, S. Molinari\inst{2}, J. C. Mottram\inst{9}, T. J. T. Moore\inst{10},  T. Onishi\inst{11}, R. Paladini\inst{1}, D. Paradis\inst{12,13}, M. Pestalozzi\inst{2}, S. Pezzuto\inst{2}, F. Piacentini\inst{14}, R. Plume\inst{15}, D. Russeil\inst{16}, F. Strafella\inst{5,6}} 
 
\institute{
Infrared Processing and Analysis Center, California Institute of Technology, Pasadena, CA, 91125
\and
INAF-IAPS, Via Fosso del Cavaliere 100, Rome, Italy 
\and
Space Telescope Science Institute, Baltimore, USA
\and
Department of Physics, Nagoya University, Japan
\and
Dipartimento di Matematica e Fisica, Universit\'a del Salento, Lecce, Italy
\and
INFN, Sezione di Lecce, Lecce, Italy
\and
Solar-terrestrial Environment Laboratory, Nagoya University, Japan
\and
National Astronomical Observatory, Japan
\and
Max Planck Institute for Astronomy, K\"onigstuhl 17, 69117 Heidelberg, Germany
\and
Astrophysics Research Institute, Liverpool John Moores University, Liverpool, UK
\and
Department of Physical Science, Osaka Prefecture University, Japan
\and
Universit\'e de Toulouse; UPS-OMP; IRAP; Toulouse, France 2 
\and
CNRS; IRAP; 9 Av. du Colonel Roche, BP 44346, F-31028, Toulouse, cedex 4, France 
\and
Departement of Physics, University of Rome La Sapienza, Piazzale Aldo Moro 5, 00185, Rome, Italy
\and
Department of Physics and Astronomy and the Institute for Space Imaging Sciences, University of Calgary, Calgary, AB T2N IN4, Canada
\and
Aix Marseille Universit\'e, CNRS, LAM (Laboratoire d'Astrophysique de Marseille), Marseille, France
}

\offprints{marcella.veneziani@stcorp.nl}

\abstract{We present the physical and evolutionary properties of prestellar and protostellar clumps in the Herschel Infrared GALactic plane survey (Hi-GAL) in two large areas centered in the Galactic plane and covering the tips of the long Galactic bar at the intersection with the spiral arms. The areas fall in the longitude ranges $19^\circ<\ell<33^\circ$ and $340^\circ<\ell<350^\circ$, while latitude is $-1^\circ< b <1^\circ$. Newly formed high mass stars and prestellar objects are identified and their properties derived and compared. A study is also presented on five giant molecular complexes at the further edge of the bar, identified through ancillary $^{12}$CO(1-0) data from the NANTEN observatory.}
{One of the goals of this analysis is assessing the role of spiral arms in the star-formation processes in the Milky Way. It is, in fact, still a matter of debate if the particular configuration of the Galactic rotation and potential at the tips of the bar can trigger star formation. }
{The star-formation rate was estimated from the quantity of proto-stars expected to form during the collapse of massive turbulent clumps into star clusters. The expected quantity of proto-stars was estimated by the possible final cluster configurations of a given initial turbulent clump. This new method was developed by applying a Monte Carlo procedure to an evolutionary model of turbulent cores and takes into account the wide multiplicity of sources produced during the collapse.} 
{The star-formation rate density values at the tips are $1.2\pm0.3\;10^{-3}\mathrm{\frac{M_\odot}{yr\;kpc^2}}$ and $1.5\pm0.3\;10^{-3}\mathrm{\frac{M_\odot}{yr\;kpc^2}}$ in the first and fourth quadrant, respectively. The same values estimated on the entire field of view, that is including the tips of the bar and background and foreground regions, are $0.9\pm0.2\;10^{-3}\mathrm{\frac{M_\odot}{yr\;kpc^2}}$ and $0.8\pm0.2\;10^{-3}\mathrm{\frac{M_\odot}{yr\;kpc^2}}$. The conversion efficiency indicates the percentage amount of material converted into stars and is approximately 0.8$\%$ in the first quadrant and 0.5$\%$ in the fourth quadrant, and does not show a significant difference in proximity of the bar. The star forming regions identified through CO contours at the further edge of the bar show star-formation rate and star-formation rate densities larger than the surrounding regions but their conversion efficiencies are comparable.}
{The tips of the bar show an enhanced star-formation rate with respect to background and foreground regions. However, the conversion efficiency shows little change across the observed fields suggesting that the star-formation activity at the bar is due to a large amount of dust and molecular material rather than being due to a triggering process.}

\authorrunning{M. Veneziani et al.}
\titlerunning{Star formation at the tips of the Galactic bar with Hi-GAL}
\maketitle

\section{Introduction}

{The {Milky Way} (MW) presents many different environmental conditions, from the very active central system constituted by the super massive black hole and the Galactic bar, to the more quiescent regions located between the arms or at the periphery of the Galaxy. This variety of physical properties makes the MW a perfect laboratory for studying star-formation processes in different natal environments and under different interactions, and to approach a model which could unify them all.}

The star-formation activity {in the central region of the {MW}, and in particular on the Galactic bar,} is not well known. This is partly due to the particular configuration of the Galactic rotation, that might alter the normal dynamics of gas and interstellar medium (ISM). Another reason is that photometric observations of the region, including the center and the Galactic bars, are biased by source confusion along the line of sight. Due to this overlap, selecting the objects located at the Galactic bar and understanding their dynamics is not trivial. According to the information about the MW structure currently available, mostly based on spectroscopic observations of the ISM and star counts (see, for example~\cite{Benjamin05,Rodriguez08}), the MW is a barred spiral Galaxy with two main arms and several secondary arms. The long in-plane Galactic bar~\citep{Lopez01,Benjamin05,Lopez07,Amores13} has {  length, horizontal and vertical thickness of approximately 7.8 kpc $\times$ 1.2 kpc $\times$ 0.2 kpc, respectively. The major axis is inclined by an angle of approximatley $42^\circ$ with respect to the Sun-Galactic center (hereafter GC) line. } It intersects the Scutum-Centaurus spiral arm in the first quadrant ($\ell\sim31^\circ$) at a distance of $\sim$6 kpc from us, and the Perseus arm in the fourth quadrant ($\ell\sim346^\circ$) at a distance of $\sim$11 kpc. The dynamics of cold molecular clouds and the way in which they collapse to form protostellar objects in those transition regions is not clear, and the role of the Galactic arms in this process is uncertain.
Gas orbiting along the edge of the bar might enhance {star formation (hereafter SF) due to a shock,} and the enhanced gravitational potential at the tips creates an aggregation of molecular clouds with a higher rate of cloud-cloud collisions than in the rest of the galaxy~\citep{Athanassoula92,Tasker09,Ohishi12}. 

{  A large quantity of far-infrared (FIR) data has been acquired at the bar edge in the first quadrant, providing a wealth of information about SF in that region. }
For example, the data on the first quadrant indicate that that region has been subject to a {  mini-starbust} process (see for example~\cite{Bally10,Nguyen11}), suggesting that the gravitational potential configuration at the intersection between the edge of the bar and the Galactic arm, might trigger SF. {  Such a scenario has not been confirmed for the far side of the bar due to the lack of data preventing a comparison between the two sides and a consistency check of the similarities or differences in their dynamical processes. In fact, this field is much less studied due also to the lower sensitivity limit in mass and luminosities caused by the larger distances of the bar in the fourth quadrant with respect to the first quadrant. Recently,} this lack of information in the fourth quadrant has been removed in the FIR  by the   Herschel/Hi-GAL (Herschel Infrared Galactic plane survey, \cite{Molinari10a,Molinari10b}) that  acquired photometric images in the PACS and SPIRE bands (70, 160, 250, 350, 500 $\mu$m) of a two-degree wide strip of the entire GP. Because of its resolution and spectral coverage, the dataset  is sensitive to the very early stages of massive star formation, from high mass starless cold molecular clouds to young protostellar clumps. Previous analysis on prestellar and protostellar objects properties observed in the far-infrared and millimeter wavelength {  with Herschel data} are reported in~\cite{Bracco11,Elia10,Elia13,Paradis10,Paladini12,Veneziani10,Veneziani13a,Veneziani13b}.

{  The Galactic bar position is still under debate,  with diverging results present in literature \citep{Englmaier1999,Hammersley10,Benjamin05,Nishiyama2005,Lopez07}. 
Data indicate that there are at least two {non-axisymmetric} structures: a Galactic bar with size 3.1$-$3.5\,kpc, orientation $\phi_{bar}$\,=\,$20^\circ-25^\circ$ and axis ratio (length-width-heigh) 10:4:3,  also refered as COBE/DIRBE bar or triaxial bulge, and a thinner long stellar bar of size $4.4\pm0.5$\,kpc, orientation $\phi_{bar}$\,=\,$44\pm10^\circ$, detected by red-clump giants and by GLIMPSE data \citep{Benjamin05,Churchwell2009}.}

In this paper, we study the star-formation processes occurring both in a very active area, specifically the edges of the Galactic bar, and in the peripheral regions of the MW along the same line of sight. 
{  We followed the MW description of~\cite{Vallee08}, where different tracers are employed, mostly HI, to derive a velocity map of the Galaxy, and therefore the tangent points and the Galactic arms position. According to this model, the beginning of the Scutum-Centaurus arm is located at $\ell\sim32^\circ$, while the beginning of the Perseus arm is at $\ell\sim339^\circ$. {We have defined the Galactic longitude coverage of our fields by increasing {the} outer edges by one degree {beyond the origin of the} the spiral arms to take into account the uncertainties in these positions.  We selected the field inner edges at Galactic longitude $\ell = 19^\circ$, allowing a tolerance of one degree with respect to the inner border of COBE/DIRBE bar reported at $\sim 20^\circ$ \citep{Englmaier1999}, and  at $\ell = 350^\circ$, longitudes where the long bar falls under {the} GLIMPSE detection limit \citep{Benjamin05}.
With this choice, we expect to cover the outer edges of the long Galactic bar, the tip of the COBE/DIRBE bar and also {portions} of spiral arms and small interarm regions.}}

We present massive prestellar and protostellar clumps properties, such as temperatures, masses, evolutionary stages and star-formation rate (SFR), in the longitude ranges $19^\circ<\ell<33^\circ$ and $340^\circ<\ell<350^\circ$. {  We selected the longitude coverage of the field in order to cover the tip of the bar and the beginning of the spiral arms. }  In this work, we first show results for the entire sample along the line of sight, and then focus on sources located at the edges of the bar to compare physical properties and activity of the star forming regions in the two sides. The star-formation rates are estimated through a protostar counts procedure, based on their masses and bolometric luminosities, adapting to the Hi-GAL dataset from the YSO evolutionary model of~\cite{McKee03} and further developed in~\cite{Molinari08}.

The paper is organized as follows: in Section~\ref{sec:data} we describe the dataset and briefly summarize the pipeline {from images to stellar properties retrieval}; in Section~\ref{sec:counts} we report the protostar counts method and the {evolutionary parameters estimate at the tips of the bar and in the background and foreground regions}; in Section~\ref{sec:bar} we present analysis on the star forming regions at the edges of the bar, focusing on the further side. Conclusions are drawn in Section~\ref{sec:conclusions}.

\section{Observations and source properties}~\label{sec:data}

\subsection{Herschel/Hi-GAL}

The Hi-GAL survey is a photometric observation of the entire GP in the PACS and SPIRE bands (70, 160, 250, 350, 500 $\mu$m) in $2^\circ\times2^\circ$ tiles.  
In this paper we make use of the final band-merged source catalog to study the global physical properties of two regions whose line of sights cross the edges of the long Galactic bar and the beginning of the spiral arms. 

The steps followed for the sky map production, photometric analysis and spectral energy distribution (SED) property extraction are thoroughly described in~\cite{Molinari15} and~\cite{Elia15}. Here we provide a brief summary. The raw dataset is converted into sky images by means of the ROMAGAL algorithm which adopts a generalized least square (GLS) minimization procedure. Since the GLS mapmaking technique is known to introduce artifacts in the maps, namely crosses and stripes {  caused by} of the brightest sources, a weighted post-processing of the GLS maps (WGLS,~\cite{Piazzo12}) has been applied to finally obtain images in which artifacts are removed or heavily attenuated.

\subsection{CO(1-0)}

In Section~\ref{sec:bar} we complement dust observations with ancillary $^{12}$CO(1-0) and $^{13}$CO(1-0) data to identify and estimate the properties of molecular gas aggregations on the edges of the Galactic bar. 
In the first quadrant, we make use of $^{13}$CO(1-0) data from the Galactic Ring Survey (GRS), collected using the SEQUOIA multi-pixel array receiver on the FCRAO 14 m telescope. The dataset {  has} an angular resolution of  46'' producing a data cube with a pixel size of 22". It covers the velocity range $-5<v_{\mathrm LSR}<135$ km/s with a resolution of 0.2 km/s. 
The antenna temperatures {  have} been converted into brightness temperatures by dividing by the factor 0.48~\citep{Jackson06}.

In the fourth quadrant, $^{12}$CO(1-0) data were collected with NANTEN, a 4 m radio telescope installed at the Las Campanas Observatory in the Atacama desert~\citep{mizuno04}, during a GP survey from 1996 to 2003. 
The NANTEN beam has a Full Width Half Maximum (FWHM) of 2.6', which is more than four times larger than the SPIRE 500$\mu$m band beam (FWHM = 35''). The data cube has a pixel size of 4' and covers the velocity range $-241<v_{\mathrm LSR}<99$ km/s with a resolution of 1 km/s. The rms fluctuations are $\sim3.2$ K and the antenna temperatures have been converted into brightness temperatures by dividing by the factor 0.89~\citep{Ogawa90}. 

\subsection{Distance determination}\label{sec:distance}

Distance determinations are particularly difficult near the Galactic bar mainly because of source crowding along the line of sight. As detailed in~\cite{Elia15}, distances have been determined from CO(1-0) spectral line observations combined with extinction maps to break the near-or-far degeneracy. Nevertheless, in case of the areas near the GC, the amount of overlapping material makes extinction measurements highly uncertain.

{  To have further confirmation of our distances, we compare the values obtained through the Hi-GAL procedure with the estimates of the Bolocam Galactic plane survey~\citep[BGPS,][]{EB13}. The two surveys overlap in the first quadrant and adopt two different methods providing independent distance determination. The BGPS derives a Bayesian distance probability density function, measuring the velocities of cold clumps through followup observations of dense gas. The near-or-far distance ambiguity is then resolved by constraining the a-priori probability density function with the mid-infrared dust absorption in molecular regions. A total of 326 BGPS sources fall in the Hi-GAL region considered in this paper in the first quadrant. We found a positional match between the two catalogs for 238 sources, within a diameter of 70", meaning twice the beam FWHM of the Bolocam 1.1 mm and the Herschel 500~$\mu$m bands. Out of these 238 matches, 167 sources (70$\%$) show distance values consistent within the error on the Hi-GAL procedure, which is estimated to be 20$\%$ of the {corresponding} distance. For the rest of the sample, we consider the possibility of a different near-or-far assignment by swapping the near with far distances and viceversa in the Hi-GAL catalog, obtaining that most of the mis-matches (20$\%$ of the initial sample) are due to a different distance ambiguity resolution. This last set of sources is randomly located within the field, suggesting that the difference in the distance ambiguity resolution between the two procedures is not due a {feature} of the region. }

\subsection{Compact source properties}

The source detection and photometry were performed by means of the CuTEx algorithm (Curvature Thresholding EXtractor, \cite{Molinari11}) which double-differentiates the sky maps and considers the curvature variations above a given threshold. The threshold varies locally according to the properties of the region, like source crowding or intensity and variability of the background, and this allows us to perform a thorough detection in difficult environments such as the GP and to collect not only point sources (sources spatially unresolved by the instrumental beam), but also compact objects which cover more than one Point Spread Function (PSF). The source profiles were then fit with a 2D elliptical Gaussian plus an underlying inclined plateau. 

The {main physical properties of a source} are estimated by fitting the SEDs with a modified {  single temperature}  black body function using a grid of models. {  The Hi-GAL clump masses have been calculated fitting a modified black body to the observed SED between 160 and 500$\mu$m~\citep{Elia13} complemented with ground-based {sub-mm} photometry where available (details will be provided in \citep{Elia15}). The simple single-temperature modified black-body model is able to well reproduce the envelope emission, but it fails for the inner core in particular when a protostar is already formed. 
For such a reason we excluded the 70 $ \mu$m emission  in the SED fitting, but it has been considered for source classification, as explained in Section~\ref{sectionclass}.  In order to have a clean sample, with very little contamination from different kind of sources, strong constraints are applied to fluxes, as explained in~\cite{Elia13}.} The associated uncertainties come mainly from background removal and calibration.  {  The SEDs of each object are built by combining Herschel measurements with 870\,$\mu$m and 1100\,$\mu$m fluxes from the ATLASGAL~\citep{Schuller09} and the BOLOCAM Galactic plane~\citep{Aguirre11} surveys, when available. }

{  As mentioned in the previous paragraph and following~\cite{Schneider12}, the fit was performed on all bands {except} the 70$\mu$m in order to avoid contamination due to the possible presence of internal protostellar activity whose emission produces an excess with respect to a single black body emission.} Source kinematic distances are estimated from CO and extinction maps with the same procedure described in~\cite{Russeil11}. 
{  Possible local motions due to the Galactic bar were not taken into account in our distance estimate. Nonetheless, the errors on the distance introduced by these effects in the sample associated to the bar {are} smaller than the near-or-far distance ambiguity discussed in Section~\ref{sec:distance}.   }
Combining SED fit results with distances we estimated bolometric luminosities and masses for each source and therefore study the evolutionary stage (see Section~\ref{sec:bar}).

\subsection{Candidate protostellar clump selection criteria}
\label{sectionclass}

The source sample is separated in three subsets after the SED fitting: {  protostellar, prestellar and unbound clumps}. A detection in the 70$\mu$m band is considered mandatory for candidate protostellar clumps. In fact, the 70$\mu$m emission has been demonstrated to be correlated with the internal luminosity of a protostar {by}~\cite{Dunham08}, and moreover we do not expect big grains in ISM clouds, only heated by the interstellar radiation field, to emit in that wavelength. As for the separation between prestellar objects and unbound clumps, we adopted the threshold in the mass vs radius diagram $M(r)>460M_\odot(r/pc)^{1.9}$ provided by~\cite{Larson81,Kauffmann2010b}. {~\cite{Lada10,Lada12} suggest that SF becomes relevant for column densities higher than $120-130\;M_\odot\;pc^{-2}$. These values correspond to a mass-radius law $M\;= \pi \times\; 120-130\;r^2\;=\;380-410\;r^2$, slightly lower than our threshold. {See also~\cite{Evans14} and references therein for more on thresholds for SF. }}

\begin{figure}[!t]
\begin{center}
\includegraphics[width=0.48\textwidth]{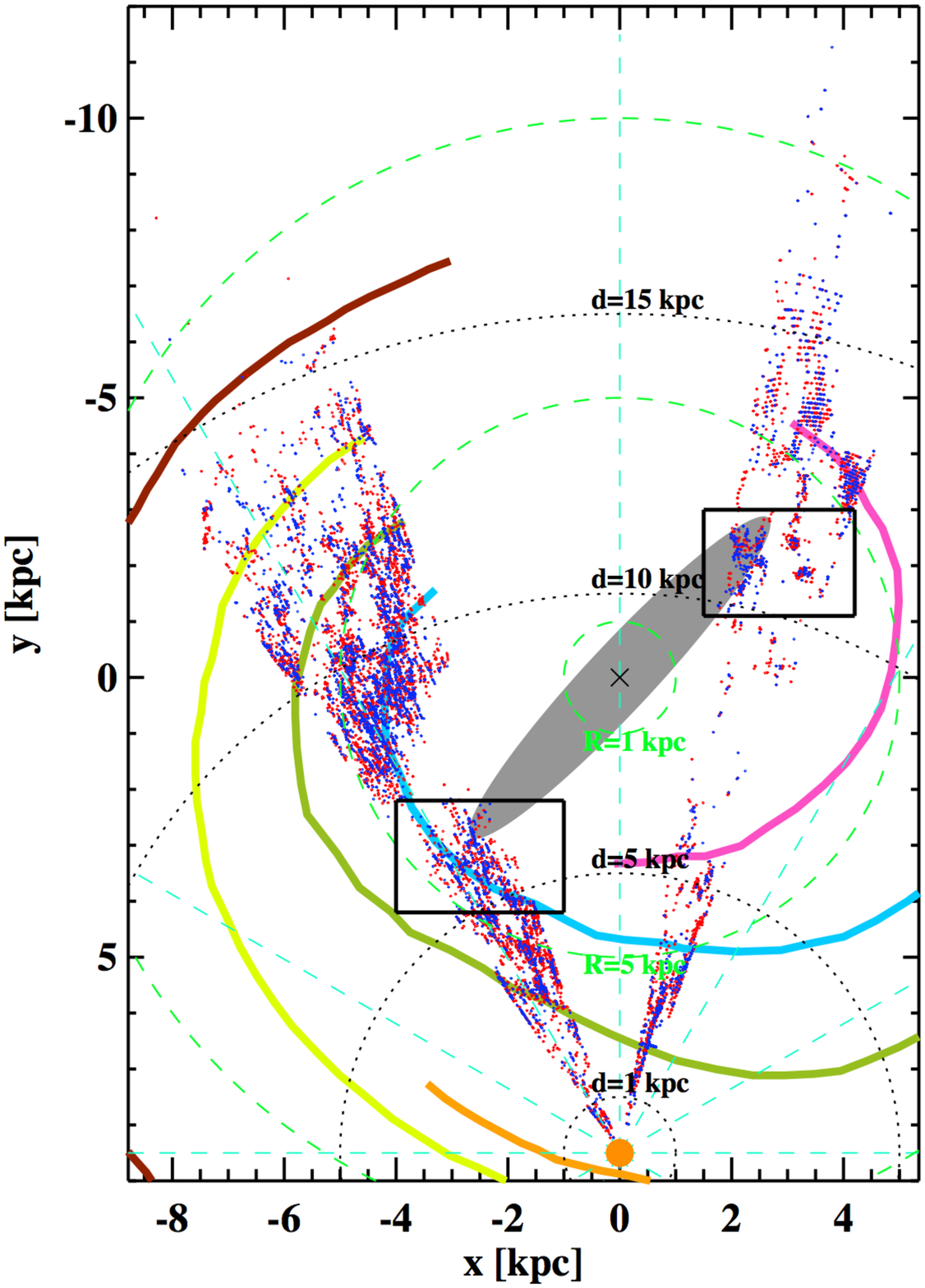}
\caption{Galactic distribution of the identified clumps in the first and fourth quadrant. {  The colored thick lines trace the empirical Galactic spiral arms as reported by~\cite{Drimmel01}.} Blue dots indicate sources with a 70$\mu$m emission (candidate protostellar sources), while red dots indicate prestellar objects. The black dotted circles indicate the heliocentric distance, while the green dashed circles indicate the galactocentric radius. The long bar as from~\cite{Amores13} is shown in gray, and the black rectangles indicate the sources considered at the edges of the bar in the present paper. The Sun is at the bottom of the image (orange circle) and the GC is represented by the black cross at the position (0,0). {  The colored curves trace the empirical position of main spiral arms mapped by HII regions and dust \citep{Russeil03,Vallee05}}}
\vspace{0.5cm}
\label{fig:galpos}
\end{center}
\end{figure}

In Figure~\ref{fig:galpos} we show the Galactic distribution of the identified objects on a schematic view of the MW~\citep{Drimmel01}. Empirical Galactic spiral arms as mapped by HII regions and dust~\citep{Russeil03,Vallee05} are also shown. 
Protostellar objects  (sources with a 70$\mu$m detection), and prestellar objects (which do not show emission in the 70$\mu$m band) are plotted.
{We have a total of 2954 protostellar objects in the first quadrant and 1778 in the fourth quadrant.
Prestellar sources are 5684 and 4365 in the first and fourth quadrant, respectively.  This corresponds to ratios of protostellar-to-prestellar objects of $\sim$ 0.52 and 0.41 for the first and fourth quadrant, respectively. {Since there is no a priori reason for these ratios to be different, the small discrepancy ($\sim$ 20\%) suggests that confusion due to the larger distances towards the fourth quadrant, might play a role when identifying suitable protostellar and prestellar candidates. We note, however, that {\cite{Enoch08}, in a study of the Perseus, Ophiucus and Serpens nearby clouds, find an approximately equal number of prestellar and protostellar cores in each cloud, while in the third quadrant, using a similar procedure as in this paper and where contamination is less of an issue, the protostellar to prestellar ratio is $\sim$ 0.37 \citep{Elia13}.}  }}

\begin{figure*}[htpb]
\centering
\makebox[\textwidth]{\includegraphics[width=.7\paperwidth]{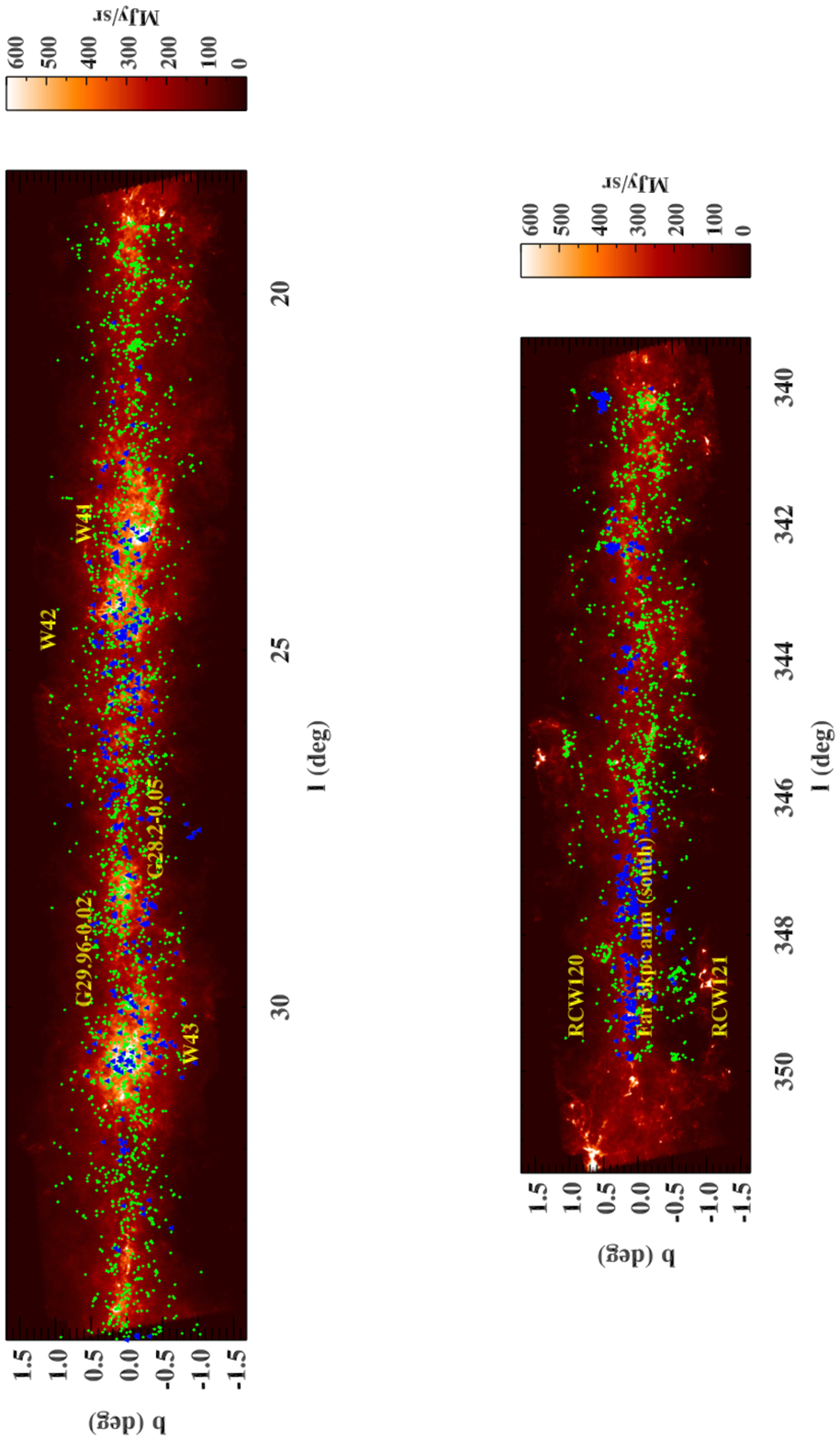}}
\caption{Images of the two observed areas in the first (top) and fourth (bottom) Galactic quadrant with Herschel/SPIRE at 500 $\mu m$. Green dots indicate protostellar objects in the entire field while blue triangles show protostellar clumps at the bar, i.e. at a distance 5$<$ d $<$ 7 kpc in the first quadrant and 10 $<$ d $<$ 12 kpc in the fourth quadrant (see Section~\ref{sec:bar} for more details). Known star forming regions from~\cite{Blum99,Leahy08,Zavagno10,Urquhart10,Sewilo11} are labeled in yellow.}
\vspace{0.5cm}
\label{fig:plw}
\end{figure*}

An overview of the two regions in the SPIRE 500$\mu$m band is shown in Figure~\ref{fig:plw}. The more famous star forming regions are indicated, and the protostellar objects in the entire line of sight and on the tips of the bar are also shown. As shown from this figure, {  our data suggest that the bar edges lie further away from the Galactic center than the regions where protostellar clumps cluster, that is between $24^\circ<\ell<31^\circ$ and $346^\circ<\ell<350^\circ$. {These clustered regions are closer to the center {than} the assumed spiral arm tangent points, respectively at $\ell\sim32^\circ$ and $\ell\sim339^\circ$. }}

\begin{figure*}[!t]
\begin{center}
\includegraphics[width=\textwidth]{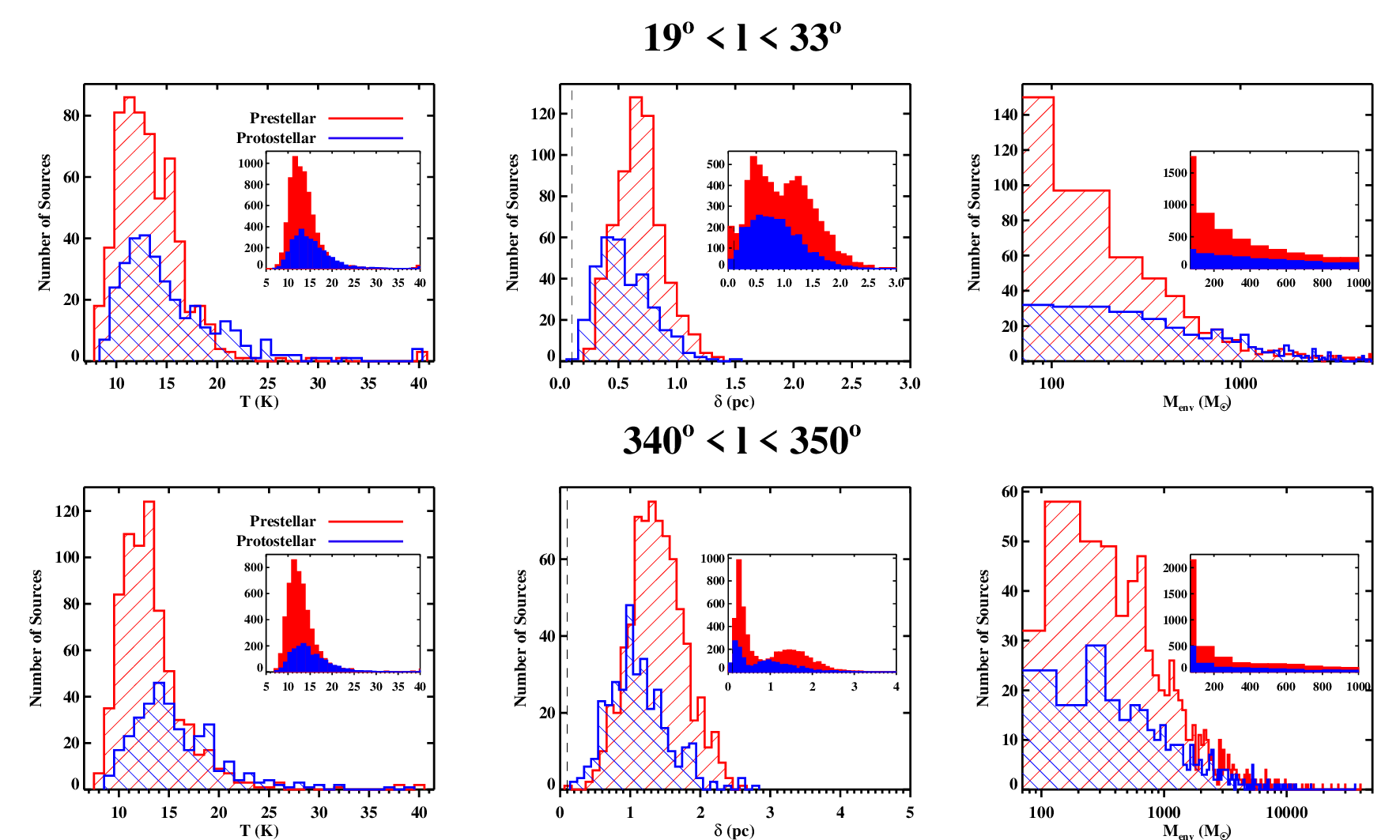}
\caption{Physical properties of prestellar (red) and protostellar (blue) sources at the tips of the bar in the first quadrant (top) and fourth quadrant (bottom). The data in the small windows show the same distributions {in the entire fields of view, therefore not only at the bar but also in background and foreground regions}. The dashed vertical line {at the very left hand side} in the size distribution marks the separation between core and clumps according to~\cite{Bergin07}. }
\vspace{0.5cm}
\label{fig:props}
\end{center}
\end{figure*}

Figure~\ref{fig:props} shows the distributions of the main physical properties of our sample of prestellar and protostellar clumps at the tips of the long bar (main windows) and in the rest of the fields (small windows) in the first (top) and fourth (bottom) Galactic quadrant. Temperatures (first column) are estimated through the SED fit. 
Sizes (second column) are the circularized FWHMs fit at {250~$\mic$} and deconvolved for the instrument beam.
We chose this band because every source, by construction, is detected at 250~$\mic$.
The {72$\%$ of sources both in the first and fourth} quadrant have radius $\delta > 0.1$~pc, {  compatible with the typical sizes of clump structures \citep[see, e.g.,][]{Williams2000} quoted by~\citet{Bergin07}}. The bimodal feature in the size distribution of sources outside the bar is due to the bimodal distribution of distances along the line of sight. 
{  Finally, envelope masses are shown in the third column. The mass distribution is likely affected by a bias towards the higher values due to completeness limits when distance becomes too large.  }
 The average values and RMS of the distributions are reported in Table~\ref{tab:props}. 

\begin{table*}[t]
\begin{center}
\space
\caption{Source distribution parameters}
\label{tab:props}
\begin{tabular}{c c c c c c c c c}
\hline
$\ell$ &  Reg &  St & \multicolumn{2}{c}{T (K)}  & \multicolumn{2}{c}{$\delta$ (pc)} &  \multicolumn{2}{c}{M$_{\mathrm {env}}$  ($10^3$M$_\odot$)}\\
(deg) & & & {average}  & rms & {average}  & rms & {average}  & rms \\
\hline
\hline
\multirow{4}{*}{19-33} &  \multirow{2}{*}{Bar}  & proto & $15.4$ & $5.3$ & $0.6$ & $0.2$ & $1.3$ & $2.2$\\
&   & pre & $13.3$ & $3.5$ & $0.7$ & $0.2$ & $0.9$ & $1.7$\\
&    \multirow{2}{*}{All} & proto & $15.4$ & $4.8$ & $0.8$ & $0.5$ & $1.7$ & $3.0$\\
&   & pre & $13.7$ & $3.9$ & $1.0$ & $0.5$ & $0.9$ & $2.2$\\
\hline
\multirow{4}{*}{340-350} &  \multirow{2}{*}{Bar}  & proto & $15.8$ & $4.6$ & $1.1$ & $0.4$ & $2.0$ & $2.9$\\
&   & pre & $13.3$ & $3.6$ & $1.4$ & $0.4$ & $1.9$ & $3.2$\\
&    \multirow{2}{*}{All} & proto & $15.3$ & $4.9$ & $0.8$ & $0.7$ & $2.1$ & $7.5$\\
&   & pre & $13.2$ & $3.9$ & $0.9$ & $0.7$ & $1.1$ & $5.8$\\
\hline
\end{tabular}
\end{center}
\footnotesize{}
\end{table*}
\space

\subsection{Mass estimate of molecular clouds}\label{sec:gas_mass}

{  Due to the different size scales, the procedure to derive large molecular cloud masses is different from the one followed for cores and clumps. Large molecular clouds extend for tens or hundreds of pc across the sky and, therefore, the estimate of their total mass from Herschel observations is highly affected by projection effects along the line of sight.
In order to reduce the projection effect we make use of the NANTEN molecular line data. In fact, different velocity channels in molecular line data probe different distances. }
The procedure we followed is the same as that described in~\cite{Elia13} and is summarized below. First, we produced an integrated CO map of each region by integrating the data cube over the velocity values of the gas in the cloud. Second, we estimated the molecular hydrogen column density in each pixel by applying the conversion factor $X$: 
\be
N(H_2) = X\;I_{CO}
,\ee
\noindent where $X = 1.4\times10^{20}\exp{(R/11\mathrm{kpc})}$  and $R$ is the galactocentric distance of the object~\citep{Nakanishi06}.
{Third, we obtained the mass by integrating $N(H_2)$ over the sky area covered by the object. The final formula is then $M_{gas}(CO) = \mu\; \mathrm{m_H}\sum_{i=1}^{N_{pix}} N(H_2)_i  A(kpc^2)_i$ where $\mu$ represents the mean molecular mass, $ \mathrm{m_H}$ is the mass of a hydrogen atom, $N_{pix}$ is the total number of pixels belonging to the cloud and $A$ is the area of each pixel expressed in kpc$^2$, once the distance is known. Adopting a relative helium abundance of $25\%$ in mass, the mean molecular mass value is $\mu=2.8$. }

\section{Clustered evolution of massive clumps and SFR estimate}\label{sec:counts}

{In this section we describe the procedure adopted to estimate the SFR of an initial clump with a given mass. 
We start from the work done by~\cite{Molinari08} that adopted the star-formation model of collapse of turbulent-supported cores near virial and pressure equilibrium {within} the harboring clump~\cite{McKee03} to describe an evolutionary sequence valid for massive objects. Their model is divided into two phases: a quick accelerating accretion phase on the central protostar followed by the cleaning up of the envelope surrounding the final star.  
With such a model it is possible to predict the mass of the final star and to estimate the formation timescales, hence the SFR, starting from measurement of envelope mass and  bolometric luminosity of  a given clump assuming that it collapses in a single high mass star. This method has been applied in previous studies of SFR from FIR/sub-mm sources ~\citep{Molinari08,Veneziani13a}. However, Herschel/Hi-GAL sensitivity and resolution mainly allow us to detect clumps that might not form single stars, but are probably the birthplaces of stellar clusters \citep{Williams2000}. Thus, to estimate a more realistic SFR from our clump sample, we {adapt the above evolutionary model to cluster-like collapse}. We follow a Monte Carlo (MC) procedure to take into account the fact that an initial clump can fragment into different final cluster configurations depending on many environmental, often unknown, factors. Cluster configurations are expected to have an initial distribution of stellar masses along the main sequence, known as stellar initial mass function (IMF). The IMF is used to indicate both the observed distribution by number of the stellar masses observed in a particular star ensemble and the theoretical probability density function of stellar masses that can be formed in a generic star ensemble \citep{Cervino13}. One remarkable aspect is that {the majority of the stellar system directly measured (field stars, open clusters, associations, and globular clusters) follows a similar IMF}. While there is still an open debate on the shape and universality on the IMF in particular at the substellar mass regime and in extreme environments, the IMF in stellar regime in our local Galactic environment appears to be universal, \citep[see][for a  review]{Bastian2010}. In this paper we use the IMF defined by Kroupa~\citep{Kroupa01} of between 0.1 and 120 M$_\odot$. The limits are given from the lower and upper values of observed stars~\citep[see, for example,][]{Andrews13}. For each MC draw, we randomly sampled the IMF as the probability density function of stellar masses to define the final cluster stellar population.}

An important parameter to set before running the MC is the {amount} of {material (gas + dust)} converted into stars. We call this parameter $\epsilon_{\star} $. {$\epsilon_{\star}$ , and it is defined as the ratio between the expected total final stellar mass and the initial mass of the object, where the initial mass is the sum of the expected final stellar mass and the mass of the envelope. According to this definition, the amount of material converted into star for a single object collapsing into one star will be }  

\be\label{eq:sfe}
\epsilon_{\star} = \frac{ m_{star}}{m_{env}+m_{star}}
,\ee

\noindent where m$_{star}$ is the mass of the final object estimated {by} digitizing the evolutionary paths of~\cite{Molinari08} and m$_{env}$ is the envelope mass of the {initial clump}. A realistic estimate of $\epsilon_\star$ is evaluated in the two regions of our Hi-GAL sample where, since we are considering several clumps at a time, the statistic is more significant. Since the number of objects is large, then $m_{env} = m_{proto}+m_{pre}+m_{unbound}$ and we find a reasonable value of $\epsilon_\star\sim1\%$. Figure~\ref{fig:sfe_dist} shows the $\epsilon_\star$ trend in the Hi-GAL region considered in this paper, as a function of the galactocentric distance (R$_{\mathrm{GAL}}$). Each data point is estimated {on the sources falling in the correspondent distance bin}.

\begin{figure}[!t]
\begin{center}
\includegraphics[width=0.5\textwidth]{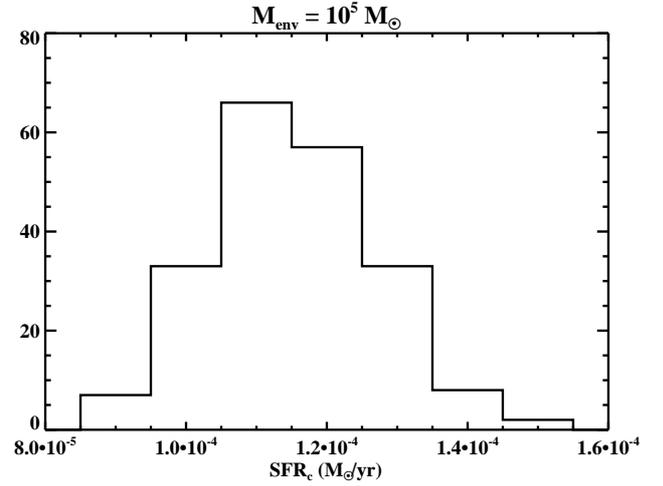}
\caption{Distribution of SFR values obtained from different cluster realizations for an initial clump mass of 10$^5$ M$_\odot$. }
\label{fig:sfr_fin}
\end{center}
\end{figure}

Given a clump with an initial mass m$_{env}$, the MC procedure for estimating its SFR with a cluster-like collapse is described as follows. For an estimate of the SFR assuming a collapse into a single high mass star we refer the reader to~\cite{Molinari08} and~\cite{Veneziani13a}. In each {run} of the MC, a random star cluster is generated by sampling the IMF. 
The selected sample is considered a realistic final configuration for the initial clump if the total bolometric luminosity of the final product is preserved, in other words, if the luminosity of the final cluster is equal to the bolometric luminosity of the final stellar mass ($L_{cluster} = L_{single}$) when assuming the clump to collapse into a single high mass star. The final luminosities have been estimated by digitizing and interpolating the values in~\cite{Molinari08} for single-star collapses. 
The SFR of the cluster configuration of the i-th {  run}  is then given by 
\be\label{eq:sfr}
{\rm SFR}_{c_i} = \sum_{k=1}^{N_\star} \frac{m_{\star_k}} {\tau_{\star_k}}
,\ee

\noindent where $N_\star$ is the number of stars in the cluster, ${m_{\star}}$ is the final stellar mass, $\tau_\star$ is the formation timescale for each star, intended as the length of time from the beginning of the collapse to when the envelope is completely cleaned up and the object is visible in the optical band. $\tau_\star$ is of the order of 10$^6$ yr. Like the luminosities, ${m_{\star}}$ and $\tau_\star$ are extrapolated from the evolutionary tracks in \cite{Molinari08}. 
We {ran} this procedure for 10,000 times for each set of m$_{env}$. The m$_{env}$ {explores the range} 140 --$10^5$ M$_\odot$, {to make sure that the values covered by our sample are fully spanned}.
The upper limit is given from the most massive source in our sample, the lower limit is given from the conventional limit of 8 M$_\odot$ for a high mass star, as according to~\cite{Molinari08} modeling a 140 M$_\odot$ clump could create an 8 M$_\odot$ star. Figure~\ref{fig:sfr_fin} shows the distribution of SFR$_{c}$ for m$_{env} = 10^5$ M$_\odot$, as an example. 

To study the variation of SFR$_c$ as a function of the efficiency, we again ran the MC, this time setting $\epsilon_\star$ to the maximum value possible to keep the condition $L_{cluster} = L_{single}$ satisfied ({  where L$_{single}$ is the bolometric luminosity of the final star when assuming a collapse into a single high mass star)}
and we call it $\epsilon_{max}$. {  The combination of bolometric luminosity preservation and very high efficiency leads to the formation of a larger number of less massive individual stars than in the previous case. This is due to the fact that these stars have a low impact on bolometric luminosity but a significant impact on the resulting total stellar mass. Therefore, the total stellar mass is larger, as expected due to the higher conversion efficiency, while the bolometric luminosity remains unchanged from before, as demanded by our constraint.} The available matter to star conversion factors in this case are  $\epsilon_{max} = [41\%, 25\%, 29\%, 11\%, 7\%]$ for clump masses of $m_{env} = [350, 700, 2000, 10^4, 10^5]$ M$_\odot$, respectively. 

\begin{figure}[!t]
\begin{center}
\includegraphics[width=0.5\textwidth]{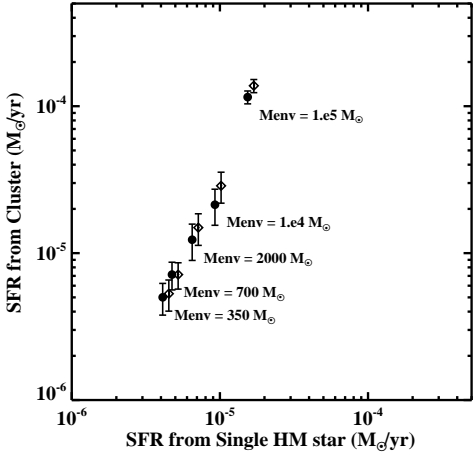}
\caption{ Comparison of SFR factors for the clump forming a high mass single star (x axis) or a cluster (y axis), assuming $\epsilon_{\star}=5\%$ (filled circles) and $\epsilon_{\star}=\epsilon_{max}$ (empty diamonds). In this last case, the values in the x axis are right shifted {by} 10$\%$, {to make the plot more clear}.}
\label{fig:sfr_fact}
\end{center}
\end{figure}

The final SFR$_c$ are reported in Table~\ref{tab:sfrfact} and the comparison of SFR$_c$ with a single star collapse (SFR$_{single}$), taken from ~\cite{Molinari08}, is shown in Figure~\ref{fig:sfr_fact}. The first column of Table~\ref{tab:sfrfact} reports the initial mass of the clump, the bolometric luminosity ($L_{cluster}$) of the final product is written in the second column. $L_{cluster}$ is the same for cluster and for a single high mass star by construction. The third and fourth columns report the SFR value assuming a cluster-like collapse with $\epsilon_\star = 5\%$ and $\epsilon_\star = \epsilon_{max}$, respectively. The SFR values assuming a single-star collapse are taken from~\cite{Molinari08} and shown in the last column for comparison. 
 
\begin{table}[t]
\begin{center}
\space
\caption{Parameters for protostellar clumps evolution}
\label{tab:sfrfact}
\begin{tabular}{c c c c c}
\hline
$m_{env}$& $L{_{cluster}}$ & SFR${_c}^{5\%}$ &  SFR${_c}^{\epsilon_{max}}$  & SFR$_{single}$   \\
 (M$_\odot$) & ($10^3$ L$_\odot$) & (M$_\odot$/Myr) & (M$_\odot$/Myr) & (M$_\odot$/Myr)  \\
\hline
\hline
350 &   25 & $5.0\pm1.2$   & $5.3\pm1.3$& 4.1\\ 
700 &   80 & $7.1\pm1.5$ & $7.1\pm1.5$&4.8 \\
2000 & 300 & $12.3\pm3.4$ & $14.9\pm3.6$& 6.5 \\
$10^4$ & 37$\cdot10^2$ & $21.4\pm5.9$& $28.7\pm6.8$&9.2 \\
$10^5$ & 12$\cdot10^4$  &$115.4\pm11.6$&$137.8\pm14.2$ &15.4 \\
\hline
\end{tabular}
\end{center}
\footnotesize{}
\end{table}
\space

As shown, when the initial clump generates a cluster instead of a single high mass star, the SFR is larger and increases with $m_{env}$. This happens because the total mass of the final product is larger, even if each star of the cluster is less massive than the high mass star produced by a single-star collapse, due to the conservation of the bolometric luminosity. 

\subsection{Star-formation rate}\label{sec:sfr}
\begin{figure}[!t]
\begin{center}
\includegraphics[width=0.45\textwidth]{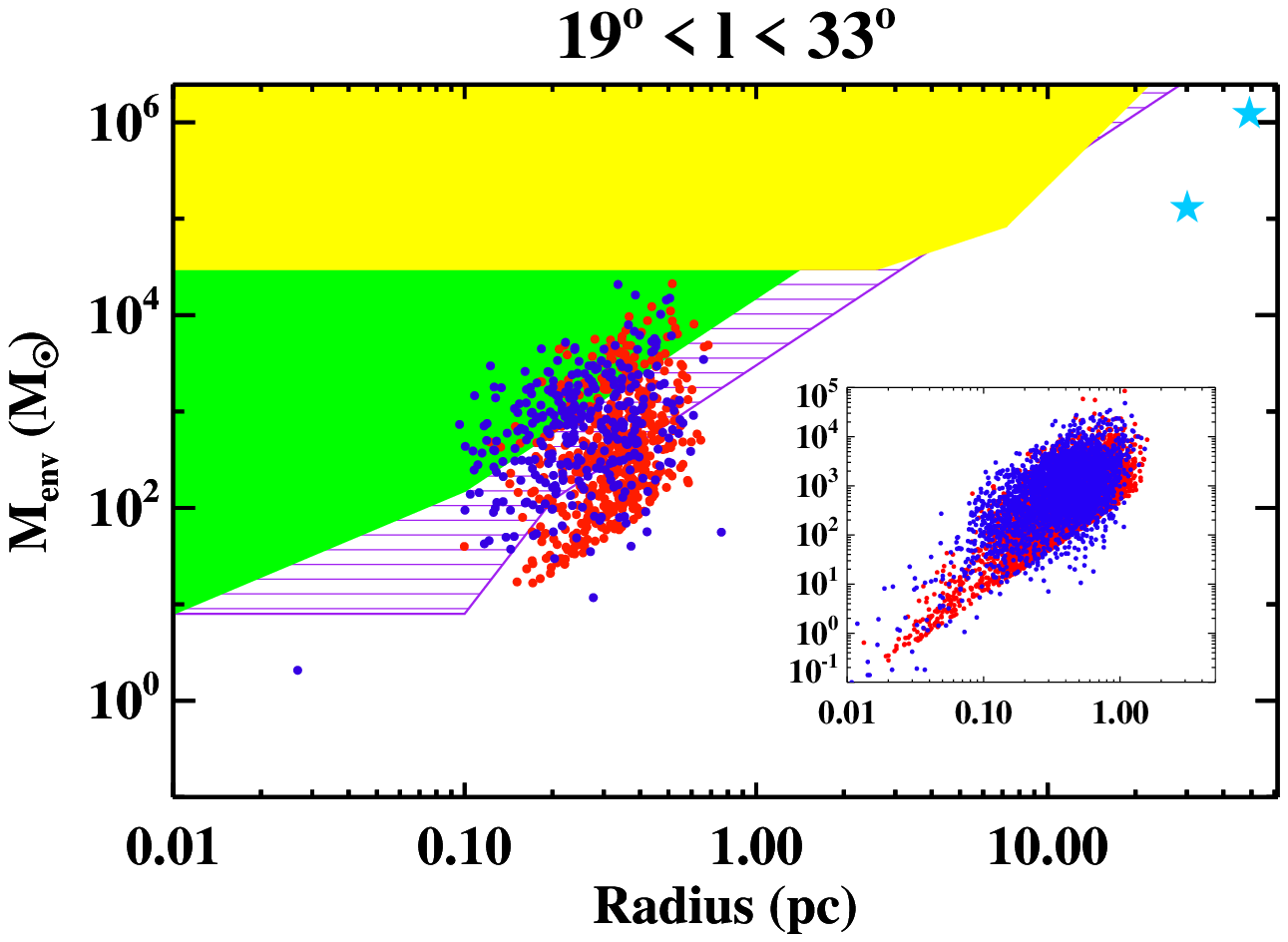}
\includegraphics[width=0.45\textwidth]{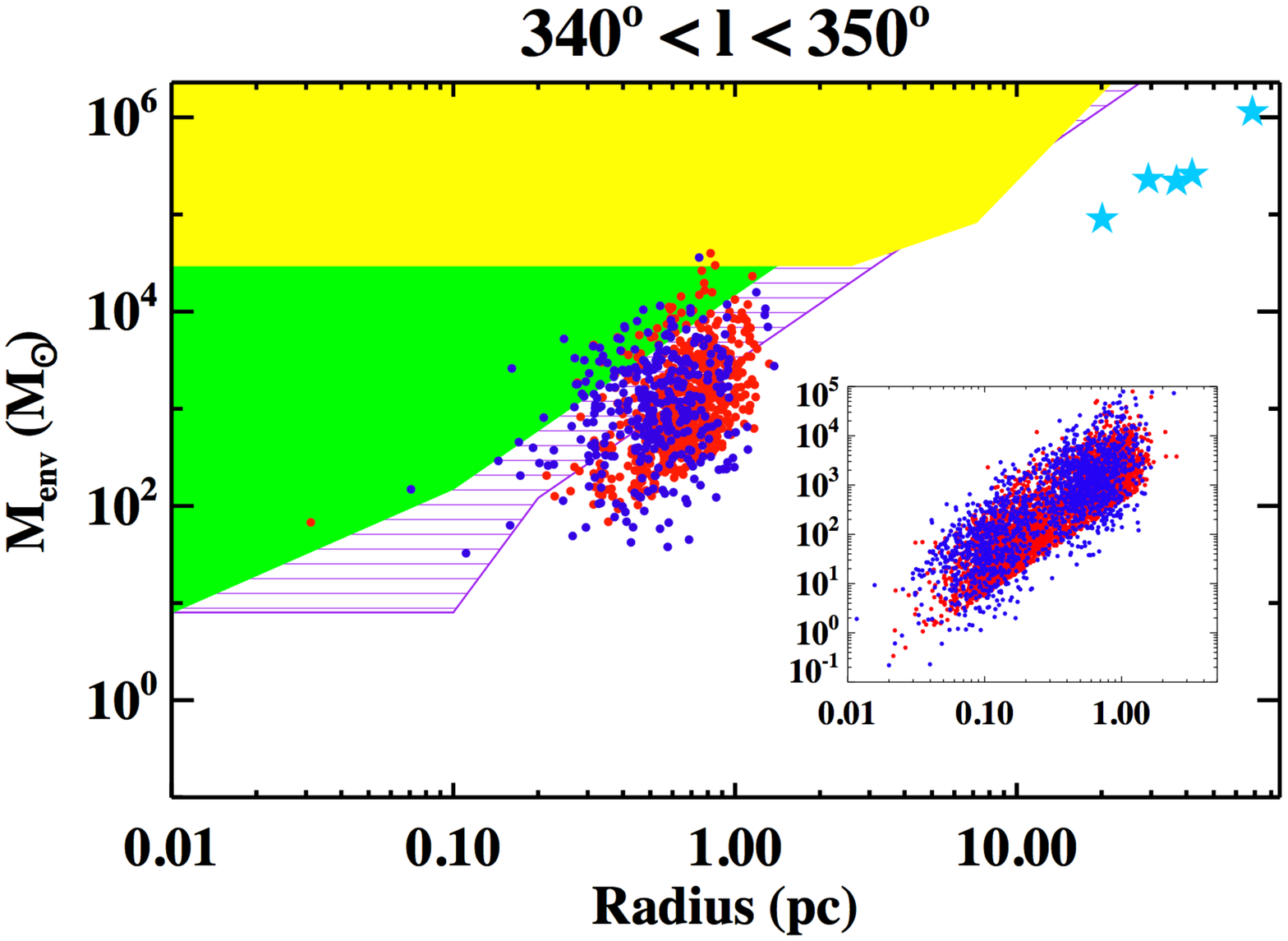}
\caption{Radius vs. envelope mass of our protostellar (blue dots), prestellar (red dots) sources at the bar. In the small windows, the same trend for the rest of the sample is also shown. 
The green solid and the purple dashed areas are the loci with medium densities higher than 1 g/cm$^2$ and 0.2 g/cm$^2$, respectively, where massive star formation starts. The yellow shaded area indicates the region of young massive clusters progenitors. {  To have a broad picture of the loci occupied by objects of a different nature, we also show the GMC (light blue stars) located in the analyzed regions, as calculated and reported in Section~\ref{sec:bar} and Table~\ref{tab:sfr_reg_4}.}}
\label{fig:masssize}
\end{center}
\end{figure}

Observations and theoretical analysis show that massive star formation within condensed molecular clumps and cores occurs when average surface densities  are higher than 1 g/cm$^2$, or about 4800 M$_\odot/$pc$^2$. This value is found in the~\cite{Plume97} sample of massive star forming regions.
~\cite{Krumholz08} also show through theoretical calculation that 1 g/cm$^2$ is the minimum density of a clump to avoid fragmentation and form massive stars. 
Nonetheless, this threshold has recently been revised  by~\cite{Butler12} who set a lower value for massive star formation of 0.2 g/cm$^2$ ($\sim$960 M$_\odot/$pc$^2$), based on GLIMPSE 8~$\mu$m high resolution observations of clumps and cores in a sample of infrared dark clouds. {A threshold lower than 1 g/cm$^2$ is also suggested by~\cite{Wu2010}, who perform a similar analysis on a sample of 50 massive clumps using different tracers. }
The relationship between mass and size of our sample of protostellar, prestellar and unbound objects is shown in Figure~\ref{fig:masssize}. In the $m_{env}$ vs R plane, sources with densities higher than 1 g/cm$^2$ and envelope masses larger than 8~M$_\odot$ occupy the green solid area in the figure. 
Clumps below the green region are going to create low mass stars and their contribution to the evolutionary parameters of the fields is discussed in Section~\ref{sec:completeness}.

In the same figure, the yellow shaded area identifies the radius-M$_{env}$ values where prestellar clumps might collapse into massive clusters, as estimated from~\cite{Bressert12}. {Massive star clusters are very rare, since most of the clusters does not remain bound after gas removal}. {They} usually form in very dense clouds with high efficiency, so that the gravity of the newborn stars counteracts the gas pressure and the group of stars remains bound.     

\begin{figure*}[!t]
\begin{center}
\includegraphics[width=0.49\textwidth]{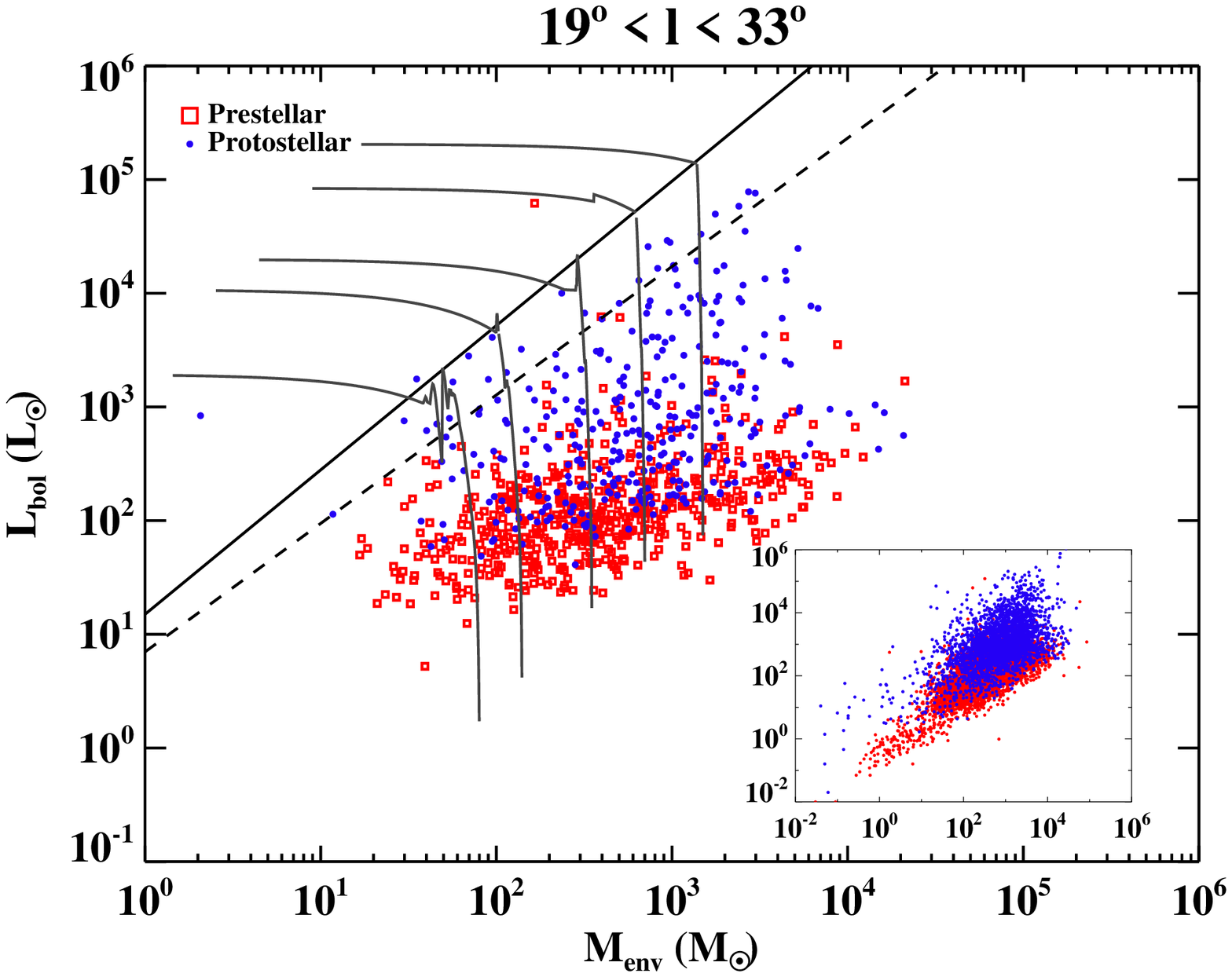}
\includegraphics[width=0.49\textwidth]{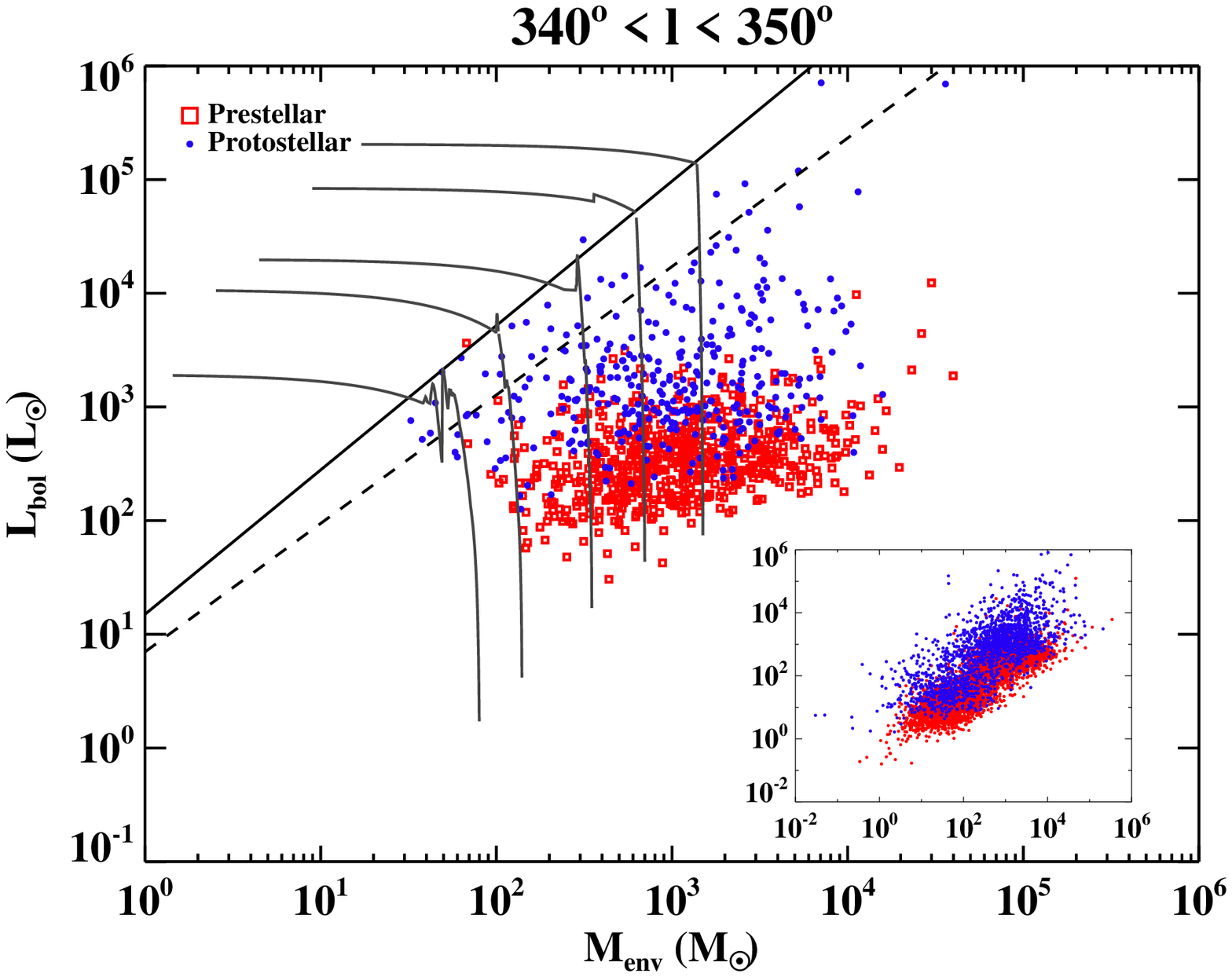}
\caption{ {Envelope mass vs. bolometric luminosity}  diagram of the sources at the bar in the first (left) and fourth (right) quadrant. The small windows show the same distribution for the rest of the sample. 
The solid and dashed lines show the best fit of high mass counterpart of Class I and Class 0 sources in~\cite{Molinari08}.}
\label{fig:lsol_msol}
\end{center}
\end{figure*}

Another qualitative analysis of the state of massive clumps can be accomplished with the $m_{env}-L_{bol}$ diagram.
In the previous sections we discussed how the combination of $m_{env}$ and $L_{bol}$ can provide the evolutionary stage of each source, as well as a prediction of the quantity of stars {with a given mass} that each protostellar clump can create. 
The $m_{env}-L_{bol}$  diagram of our sources, both prestellar and protostellar, is shown in Figure~\ref{fig:lsol_msol}. As for the previous plots, the main window shows sources located at the bar and the small window shows the rest of the sample. 
{The best log-log fit of the high mass sources counterpart of the low mass Class I sources regime is indicated, as is the high mass counterpart of the best log-log fit of the low mass Class 0 regime~\citep{Molinari08}. The evolutionary tracks each clump is supposed to follow during the collapse, according to~\cite{McKee03}, are also shown. Almost all our sources, even the protostellar clumps, appear to be at a very early stage, which can be classified as a high mass counterpart of the low mass Class 0. }

\begin{figure*}[!t]
\begin{center}
\includegraphics[width=\textwidth, angle=0]{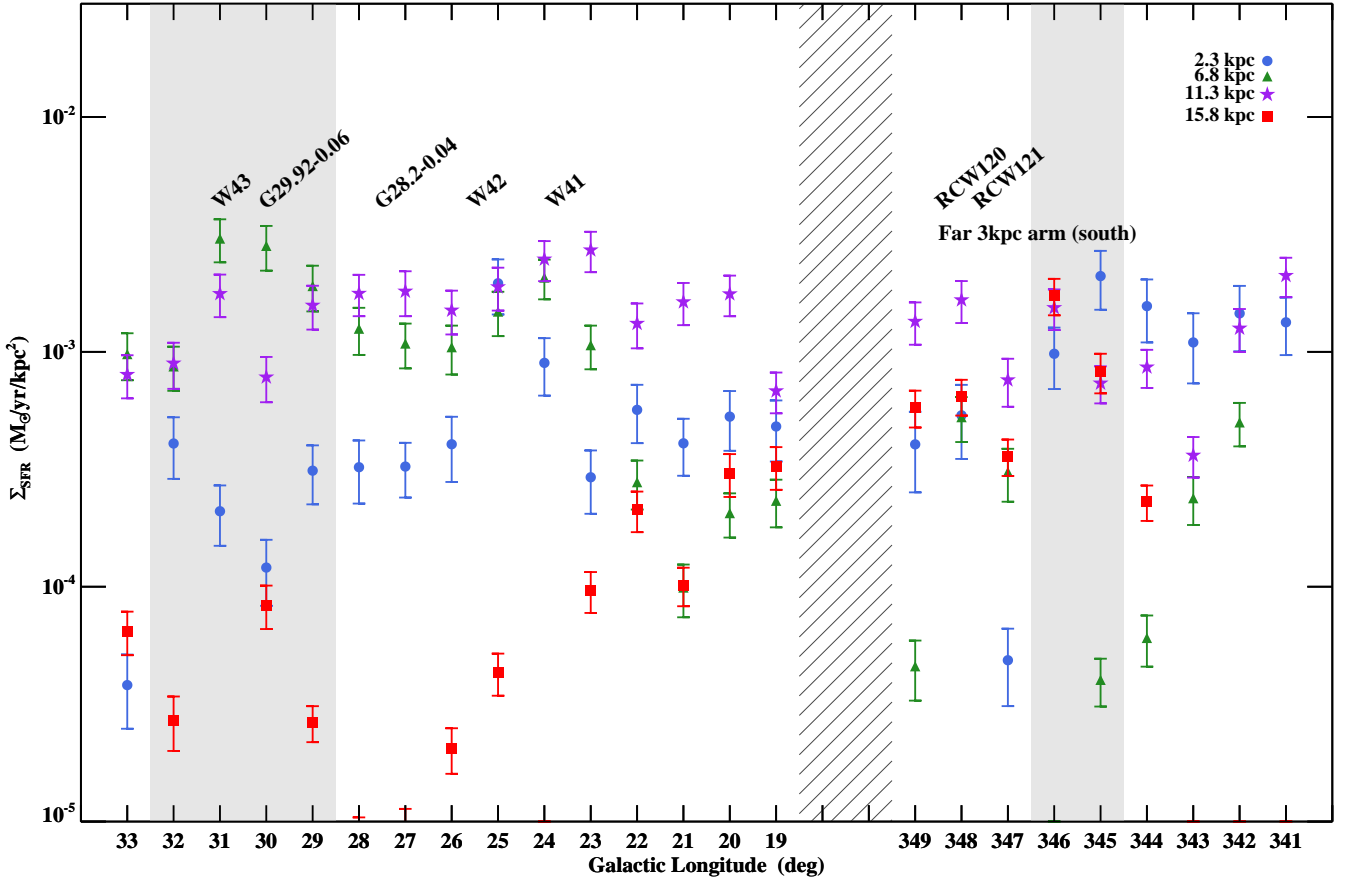}
\caption{ {Star formation rate surface density}, $\Sigma_{\mathrm  SFR}$, as a function of the Galactic longitude {in heliocentric} distance {bins}. The shaded {gray dashed region} indicates the GC, which is not covered in the present analysis, while the gray filled regions indicate the edges of the bar. Known star forming regions are also reported.  } 
\label{fig:sfr_gl}
\end{center}
\end{figure*}

We study the star-formation rate surface density ($\Sigma_{\mathrm SFR}$) as a function of the Galactic longitude (Figure~\ref{fig:sfr_gl}). {We split our sources sample in four bins of heliocentric distance centered on [2.3, 6.8, 11.3, 15.8]\,kpc and estimate the SFR from the number of protostars falling in each longitude-distance bin. Then, we normalized the SFR by the area of the Galactic sector, imaging the Galaxy as a disk viewed from above, computing the star-formation rate surface density ( $\Sigma_{\mathrm SFR}$). The result is shown in Figure~\ref{fig:sfr_gl} where each point represents the $\Sigma_{\mathrm SFR}$ of one degree longitude field for different  heliocentric distance bin. Such a representation allows {us} to highlight star-formation enhancements due to local activity. In fact, in $22^\circ\lesssim\ell\lesssim32^\circ$, $\Sigma_{\mathrm  SFR}$ is higher in the two central bins, centered on 6.8 and 11.3 kpc. The first of these bins  which contains the star forming regions W43 ($\sim6.2$ kpc), G29.96-0.02 ($\sim6.2$ kpc) and  G28.2-0.04 ($\sim5.7$ kpc) located approximately at the edge of the bar, while the second includes all the sources associated with the Perseus arm ($\sim 10$ kpc). In the closest distance bin there are  two peaks which correspond to the regions W41 ($\sim4$ kpc) and W42 ($\sim3.7$ kpc)}. The {  lower values of $\Sigma_{\mathrm SFR}$ in the outer bin, centered at 15.8 kpc, where the line of sight intersect the outer Scutum-Centaurus arm, are probably due to sensitivity limits}. Moreover, according to~\cite{Dame11}, this arm also started to move off the GP, {  consistent with the flaring and warping of the outer HI disk \citep{levine06}.}

In the fourth quadrant, $\Sigma_{\mathrm  SFR}$ is high in the first distance bin between $341^\circ\lesssim\ell\lesssim346^\circ$, due to local star forming regions, and in the third bin at $\ell \simeq 348^\circ,\;346^\circ$, due to the two main star forming regions located at the edges of the Galactic bar (see Section~\ref{sec:bar} for more details).

\begin{figure}[!t]
\begin{center}
\includegraphics[width=0.49\textwidth,angle=0]{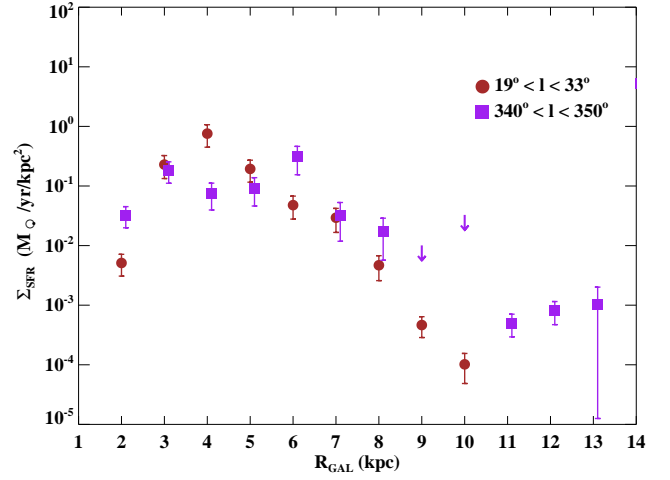}
\caption{{Star formation rate surface density}, $\Sigma_{\mathrm  SFR}$, of the two areas as a function of the Galactocentric distance (R$_{\mathrm{GAL}}$). The values of the fourth quadrant (purple squares) are slightly shifted on the X axis for better readiness.}
\label{fig:sfr_dist}
\end{center}
\end{figure}

Figure~\ref{fig:sfr_dist} reports the $\Sigma_{\mathrm  SFR}$ as a function of the galactocentric distance (R$_{\mathrm{GAL}}$). As expected, the maximum of the activity occurs within the inner 6-7 kpc ring where most of the Galactic HII regions are located, and then decreases towards the outer Galaxy. 
{The two quadrants show slightly different trends, due to the intersections with the Galactic arms. The first quadrant has its maximum of the SFR activity at  R$_{\mathrm{GAL}}\simeq4$ kpc and then decreases smoothly towards the outer galaxy, showing no significant activity consistently to~\cite{Urquhart14}. The fourth quadrant shows two maxima of the SFR activity at R$_{\mathrm{GAL}}\simeq3$ kpc and R$_{\mathrm{GAL}}\simeq6$ kpc and then reports a milder decrement with the distance, due to a larger number of protostars detected as far as 15 kpc from the GC. A similar trend is shown in~\cite{Kennicutt2012} averaged on the whole MW. The two trends, the one on the whole MW and the one shown in this paper, (i.e., focused on the tips of the bar), are consistent at  R$_{\mathrm{GAL}}\lesssim8$ kpc. At further distances, the line of sight pointing at the tips of the bar doesn't show any other significant star-formation activity. }  

In the first part of Table~\ref{tab:result} are reported the overall values for protostellar sources located at the distance of the bar (bar only) and on the rest of the fields (all).

\begin{table*}[!htb]
\begin{center}
\space
\caption{Evolutionary parameters}
\label{tab:result}
\begin{tabular}{l c c c c c}
\hline
\\
$\ell$ &  $\Sigma_{\mathrm{proto}}$ & $\Sigma_{\mathrm{SFR}}$ & $\Sigma_{\mathrm {SFR}\_LM}$ & $\Sigma_{\mathrm {SFR}\_IR}$ &$\epsilon_\star$\\
(deg) & ($\frac{1}{kpc^2}$) & ($10^{-3 }$ $\frac{\mathrm M_\odot}{\mathrm yr\cdot kpc^2}$) & ($10^{-3 }$ $\frac{\mathrm M_\odot}{\mathrm yr\cdot kpc^2}$)  & ($10^{-3 }$ $\frac{\mathrm M_\odot}{\mathrm yr\cdot kpc^2}$) & ($\%$)\\
\hline
\hline
{$19-33$} (all) &  72  &$0.9\pm0.2$  &  0.1&  ${1.0\pm0.2}$ & 0.7\\
{$340-350$} (all) &  723   &  $0.8\pm0.2$&  0.1& ${ 0.4\pm0.1}$ &0.4 \\
\hline
{$19-33$} (bar only) & 111  &  $1.2\pm0.3$& 0.2 & $0.5\pm0.2$ &  0.8  \\
{$340-350$} (bar only)  & 111 &   $1.5\pm0.3$& 0.1 & $1.4\pm0.4$ &0.5 \\
\hline
\hline
\\
& \multicolumn{3}{c}{Distance switch: near $\Rightarrow$ far}\\
\hline
{$19-33$} (all) & 72    & $0.9\pm0.2$  & - & - &0.7\\
{$340-350$} (all) &   59  &  $0.6\pm0.1$  & - & - &0.4 \\
\hline
{$19-33$} (bar only) &   9  & $1.2\pm0.3$ & - & -& 0.8  \\
{$340-350$} (bar only)  & 14  &  $1.2\pm0.3$ & - & -& 0.4 \\
\hline
\\
& \multicolumn{3}{c}{Distance switch: far $\Rightarrow$ near}\\
\hline
{$19-33$} (all) & 45    & $0.5\pm0.1$  & - & -& 0.7\\
{$340-350$} (all) &   71  &  $0.8\pm0.2$  & - & -& 0.4 \\
\hline
{$19-33$} (bar only) &   35  & $5.5\pm1.2$ & - & -& 0.7  \\
{$340-350$} (bar only)  & 1  &  $0.13\pm0.03$ & - & -& 0.6 \\
\hline
\end{tabular}
\end{center}
\footnotesize{Values of density of protostellar clumps (second column), $\Sigma_{\mathrm  SFR}$ (third column), missing $\Sigma_{\mathrm  SFR\_LM}$ due to completeness limits (fourth column), {SFR density calculated through IR luminosity as in~\cite{Li10}} and the amount of {  material} converted into stars (sixth column), in and outside the Galactic bar. The three sets of rows separated by the double horizontal line report those values for the distance configuration provided in the original catalog (first set), when switching all the distances assigned as "near" into "far" (second set), and "far" into "near" (third set), as described in Section~\ref{sec:distance}. }
\end{table*}
\space

\subsection{SFR from FIR estimator}

{Another method which has been recently adopted~\citep{Veneziani13a,Elia13} to estimate the SFR from a FIR dataset, is a monochromatic SFR indicator estimated on sub-Galactic star forming regions detected in the 70$\mic$ band of the Spitzer Space Telescope~\citep{Li10}. By comparing the SFR obtained with an independent estimator, H$\alpha$+24$\mic$ with the 70$\mic$ luminosity in those regions,~\cite{Li10} find the following SFR indicator: ${\mathrm{SFR} \left[{\mathrm M_\odot/\mathrm yr}\right]} = \frac{\mathrm L (70)}{1.067\;(\pm\;0.017)\;\times\;10^{43}\;\left[\mathrm{erg}\cdot s^{-1} \right]}$.}

{Therefore, following~\cite{Veneziani13a}, we also compute the SFR density at the tips of the bar by converting the 70$\mic$ luminosity using the same monochromatic estimator at 70$\mic$. The resulting values at the edges of the bar are $0.5\times10^{-3}\frac{\mathrm M_\odot}{\mathrm yr\cdot kpc^2}$ and $1.4\times10^{-3}\frac{\mathrm M_\odot}{\mathrm yr\cdot kpc^2}$ in the first and fourth quadrant, respectively. {As shown in Table~\ref{tab:result}}, the two methods, source counts and IR estimator, provide SFR values within the same order of magnitude, being fully consistent in the fourth quadrant at the bar, and in the first quadrant in the whole field. We report the full set of results based on 70$\mic$ luminosity, both at the edges of the bar and in the full fields, in Table~\ref{tab:result} to allow an easy comparison. }

\subsection{Conversion efficiency}
\label{sec:sfe}

In Section~\ref{sec:counts} we introduced the parameter $\epsilon_\star$, which provides information about the amount of {material} turned into stars. While a high SFR can simply depend on the amount of molecular material present in the area, variations of $\epsilon_\star$ across the MW can answer the question about the role of the bar in cloud collapse. In fact, the bar might only gather GMC and ISM due to the gravitational potential {configuration} but not trigger star formation. In this case the SFR {at the bar} is higher than in {a star forming region in the arms}, due to the larger amount of material, but $\epsilon_\star$ stays the same. On the other hand, the {bar} might trigger star formation, for example through shocks or simply enhancing the probability of cloud-cloud collisions. In this last case, $\epsilon_\star$ in the {arm} regions should be smaller than at the {tips of the bar}.

\begin{figure}[!t]
\begin{center}
\includegraphics[width=0.49\textwidth, angle=0]{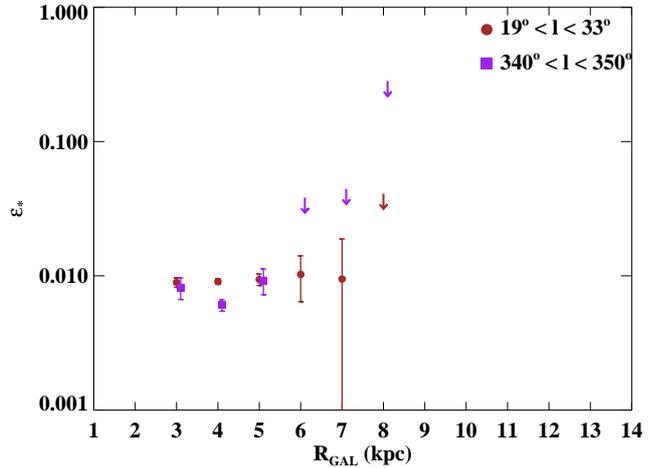}
\caption{{Conversion efficiency}, $\epsilon_\star$, as a function of the Galactocentric distance.  }
\label{fig:sfe_dist}
\end{center}
\end{figure}

Results of $\epsilon_\star$ are shown in Figure~\ref{fig:sfe_dist} as a function of the galactocentric distance and reported in Table~\ref{tab:result}, {considering for each field either the case of all the sources found there or restricting the sample to sources associated with the bar}. The error bars are estimated assuming a Poisson distribution of the number of sources falling in each distance bin.
The two samples are plotted together to determine if there is any symmetry between the two sides of the GC. Similar behavior
of $\epsilon_\star$ is seen on the two sides {of the bar, with a value around 1$\%$ for galactocentric distances between 2 and 4 kpc}. At larger distances, the first quadrant remains constant up to  8 kpc where the number of sources in each bin is not great enough to determine $\epsilon_\star$. Due to the larger heliocentric distance, $\epsilon_\star$ becomes undetermined slightly closer to the GC in the fourth quadrant, around 6 kpc. 

It is worth noticing that at the intersection of the bar with the arms, located at a galactocentric distance of $\sim4$ kpc, the first quadrants shows a value consistent with the other regions and the fourth quadrant a value slightly lower. {This seems to rule out the hypothesis of triggering at the intersection between the bar and the arms, indicating that the higher rate of star formation in the inner 7 kpc ring is due to the higher amount of molecular material, as found by~\cite{Moore12}. }

\subsection{Clump formation efficiency}

Since most stars form from clumps collapsing into clusters, the star-formation processes are strictly related to the clump formation processes and efficiency. The clump formation efficiency (CFE) describes the amount of molecular gas {  in } the GMC grouped into regions with higher densities which in turn, under certain conditions, might themselves host several cores indicating an ongoing gravitational collapse. 
According to this scenario and following~\cite{Elia13}, the CFE is defined as

\be\label{eq:cfe}
{\mathrm CFE} = \frac{m_{clump}}{m_{clump}+m_{cloud}}
,\ee

\noindent where $m_{clump}$ is, as before, the envelope mass of each clump, both prestellar and protostellar and $m_{cloud}$ is the mass of molecular cloud hosting the considered clumps,{  estimated from CO as reported in Section~\ref{sec:gas_mass}}. We adopted this procedure to estimate the CFE of large aggregations of molecular gas observed in the fourth quadrant, as reported in Table~\ref{tab:sfr_reg_4}.

\subsection{Effect of distance errors on SFR values}

{The good agreement between Hi-GAL and BOLOCAM BGPS distances leads to a good agreement also in the SFR values. The two SFR calculated with BOLOCAM and Hi-GAL distances on the 326 matching sources (see Section~\ref{sec:distance}), differ only by a few percent and are fully consistent within the errors. Their values are SFR$_{\mathrm BGPS} = 0.023\pm0.009$ and  SFR$_{\mathrm Hi-GAL} = 0.031\pm0.012$, showing that the small discrepancy does not affect the scientific results}.

{Despite the good agreement between Hi-GAL and BGPS distances and corresponding SFR values, we further} test the robustness of our results and the error on SFR and $\epsilon_\star$ estimates generated from an incorrect near-or-far distance assignment. In order to do that, we replicated the procedures adopted to estimate $\Sigma_{\mathrm SFR}$ and $\epsilon_\star$ by assigning the far distance to the sources identified as near through the extinction criterium, and vice-versa. 
45$\%$ and on 33$\%$ are the amount of protostars in the first and fourth quadrant, respectively, showing no absorption feature in the extinction maps. In this case, the source is usually assigned a far distance, since a lot of material is assumed to lie in the foreground, hiding the feature. However, {this assumption might be wrong as there is always the possibility that the source is } located in the close distance, but still larger than $\sim3$ kpc of a very crowded region, where the overlapping material would not be enough to show the absorption. 

We have then estimated the SFR swapping the near and far distances and we obtained the values reported in the second and third part of Table~\ref{tab:result}. Both areas as a whole do not seem to be affected by the distance uncertainty within the error bars. The only {significant exception are the edges of the bar when all far distances are replaced with the near. In this case, $\Sigma_{\mathrm SFR}$ increases significantly in the first quadrant and drops in the fourth quadrant, keeping $\epsilon_\star$ unchanged. These differences are due to the protostar density variation (second column of Table~\ref{tab:result}). The most affected area is the {tip} of the first quadrant, where our test shows a {$\Sigma_{\mathrm SFR}$ } increment of about a factor of 50. Despite the fact that a completely wrong near-or-far distance assignment is highly unlikely, our test shows that the first tip of the bar would be the most affected area in case most of the near sources are wrongly assigned. }

\subsection{Completeness}\label{sec:completeness}
One of the most difficult challenges in the analysis is the bias estimate due to the survey completeness limit. The more distant or fainter objects are more difficult to detect, {  hence it is expected that the sources on the edge of the bar in the fourth quadrant would be harder to detect. However, capability of {source identification} is also limited by the overall emission in the field and its crowdedness. In the case of low crowded field with an overall shallower background emission it is possible to recover fainter and more distant objects. In general, it is estimated that Hi-GAL survey is mostly sensitive to high mass clumps, thus our current analysis might miss the contribution from low mass stars. In this section, we derive the bias estimate following the same procedure described in~\cite{Veneziani13a}.}

\cite{Molinari15} ran a set of simulations where they put synthetic sources in the real sky map and recover them with the current pipeline. The amount of sources recovered provides information about the flux below which, either because the source is too faint or because of source crowding, the algorithm {does} not detect the right number of input clumps. 
{In order to estimate the completeness limit on the envelope masses of the clumps, we use the completeness limit fluxes of the 350 $\mu$m band, which are are 6.89 Jy in the first quadrant and 5.77 Jy in the fourth quadrant. } These values are converted to clump mass by assigning, as distance, the average distance of the fields. The {lower} {clump mass limits} are then 546 M$_\odot$ in the first quadrant (at $\sim$ 7 Kpc) and 400 M$_\odot$ in the fourth quadrant (at $\sim$ 12 Kpc). The corresponding {stellar mass limits} are estimated by extrapolating the {clump mass limits} with the evolutionary tracks, obtaining 16.9 M$_\odot$ in the first quadrant and 14.4 M$_\odot$ in the fourth quadrant. The number of stars with mass smaller than the stellar mass limits and down to 0.1 M$_\odot$ is then extrapolated from the high to the low mass regime by means of the Kroupa IMF and then converted into an estimate of missing SFR by applying the~\cite{Lada10} equation, (SFR = N$_{low}\frac{0.5 \mathrm M_\odot}{2 \mathrm {Myr}}$, where N$_{low}$  is the number of missing stars). {The above expression is based on the analysis of the nearby molecular clouds~\citep{Evans09} where low mass protostars and young stellar objects with that mean age are the dominant population.}

Results of this procedure can be found in the fourth column of Table~\ref{tab:result} ($\Sigma_{\mathrm SFR\_LM}$, first set of rows) inside and outside the Galactic bar. The values show that the $\Sigma_{\mathrm SFR}$ is underestimated $\sim$10$\%$ in the first quadrant and by less than $10\%$ in the fourth quadrant, due to the completeness limit.

\section{SFR at the tips of the bar}\label{sec:bar}

In this section we analyze Hi-GAL data focusing on the protostellar clumps located at the tips of the bar, to find out {whether} the two sides show similar star-formation activity. As before, we consider all clumps with a distance 5 $< d <$ 7 kpc in the first quadrant and 10 $< d <$ 12 kpc in the fourth quadrant as belonging to the {ends of the} bar. The sources satisfying these criteria are shown in Figure~\ref{fig:plw}.

In Table~\ref{tab:result} we report $\Sigma_{SFR}$ estimated on sources at the two sides of the bar  (labeled "bar only" in the Table) together with the values on the whole field of views (labeled "all"), for comparison. The values at the bar are slightly higher than the ones averaged over the full regions, especially in the fourth quadrant, while $\epsilon_\star$ has consistent values over the entire sample, again ruling out{} the hypothesis of some triggering process.

Mainly because of its location, the closest side of the bar ($29^\circ\lesssim \ell \lesssim32^\circ$) is {well} observed and studied~\citep[see for example,][]{Moore12}. The main SF region in that area is W43~\citep[see for example,][]{Blum99,Bally10,Nguyen11}, which in terms of SFR and $\epsilon_\star$ has been classified as a mini-starbust  and is located at the intersection between the bar and the Scutum-Centaurus spiral arm. Other identified star forming regions related to the bar are G23.44-0.18~\citep{Ohishi12}, at the intersection with the Norma arm, and G29.96-0.02~\citep{Beltran13} which is sometimes associated to the W43 complex. All these regions have an heliocentric distance of $\sim6.2$ kpc.
In the next section, {we make use of CO contours to identify SF regions in the fourth quadrant and estimate the physical parameters with Hi-GAL data.} 

\subsection{Star forming regions in the 4$^{th}$ quadrant}\label{sec:q4}

In this section we combine Hi-GAL far-infrared observations with molecular data, in order to better study the distribution of ISM and molecular gas in the further side of the bar. The goal of this analysis is to understand if the two tips of the bar undergo similar physical processes and if those processes are different to the ones occurring in the foreground and background. 

The further side of the bar is less observed and studied than the one in the first quadrant for several reasons. First, the distance determinations are less accurate due to confusion along the line of sight. Second, due to the position of the bar with respect to the Sun-GC line, the completeness limit in the fourth quadrant is {higher} and a large number of sources {in} the GP with $\ell > 340^\circ$ are hidden to our sight. {Third, due to the crowded line of sight and the difficulties in breaking the near-or-far degeneracy in distance determination, the position of the Galactic arms in that direction is still uncertain~\citep{Vallee08}. }

\begin{figure*}[!t]
\begin{center}
\includegraphics[width=0.45\textwidth,angle=0]{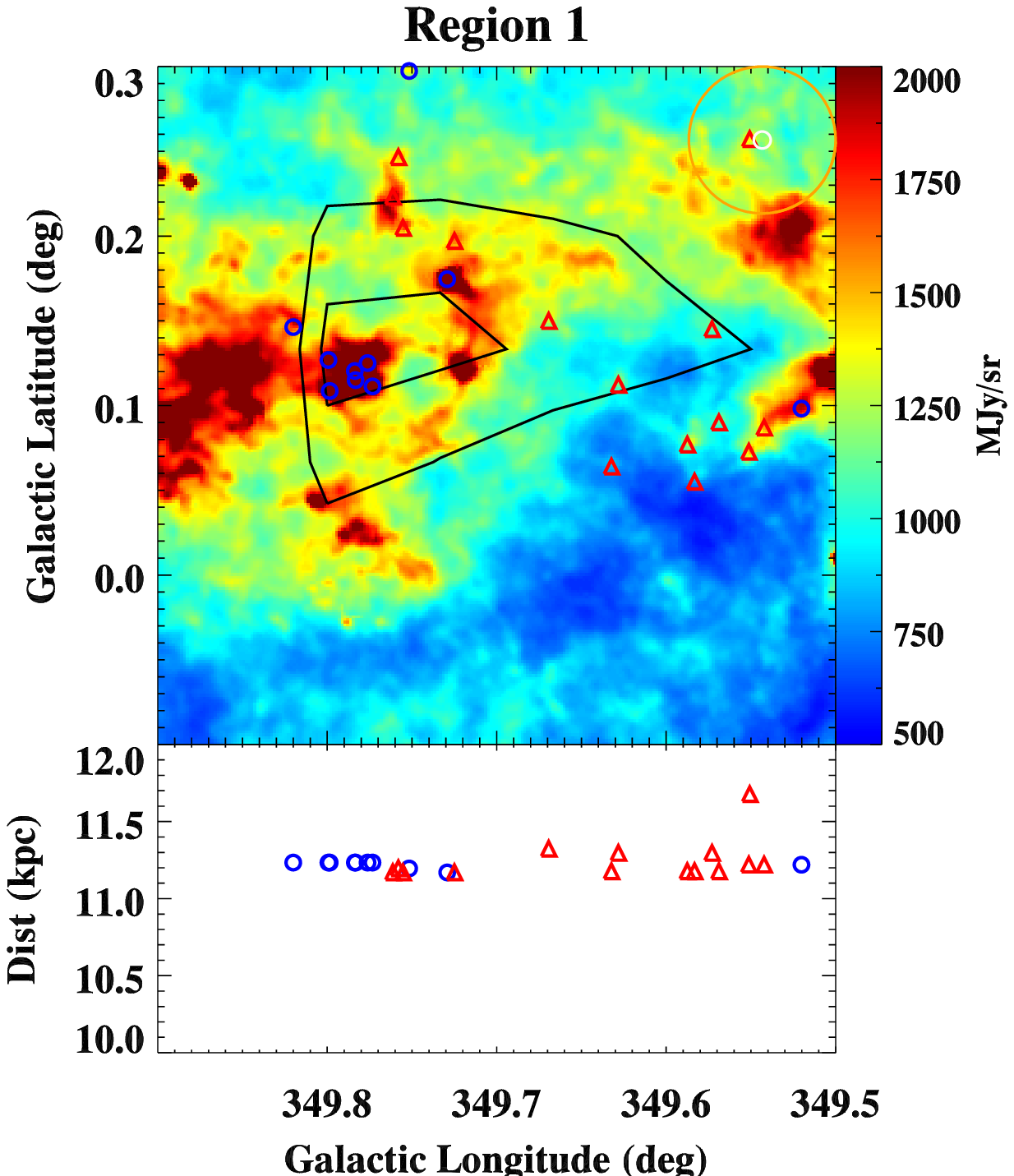}
\includegraphics[width=0.50\textwidth,angle=0]{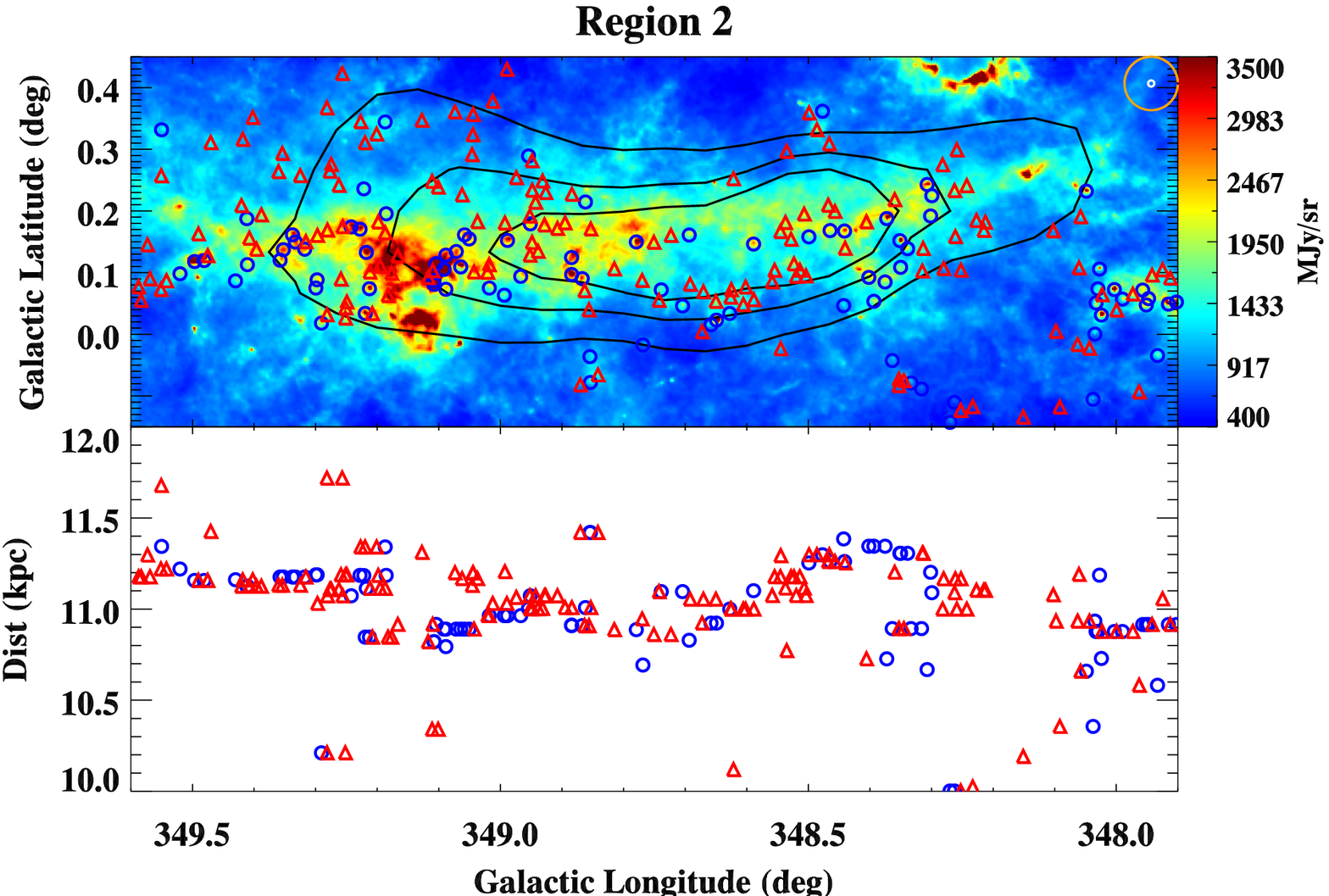}
\includegraphics[width=0.47\textwidth,angle=0]{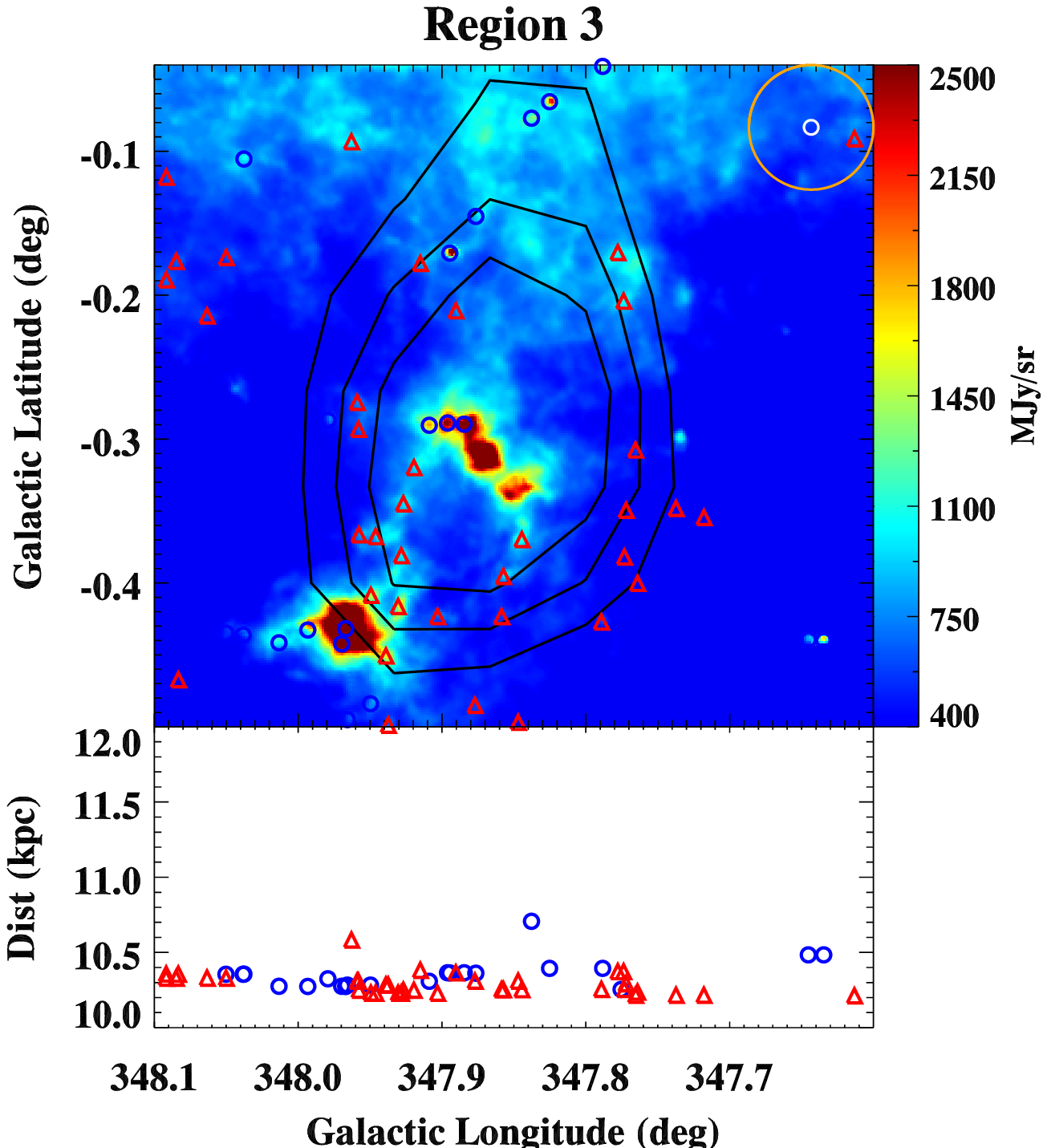}
\includegraphics[width=0.48\textwidth,angle=0]{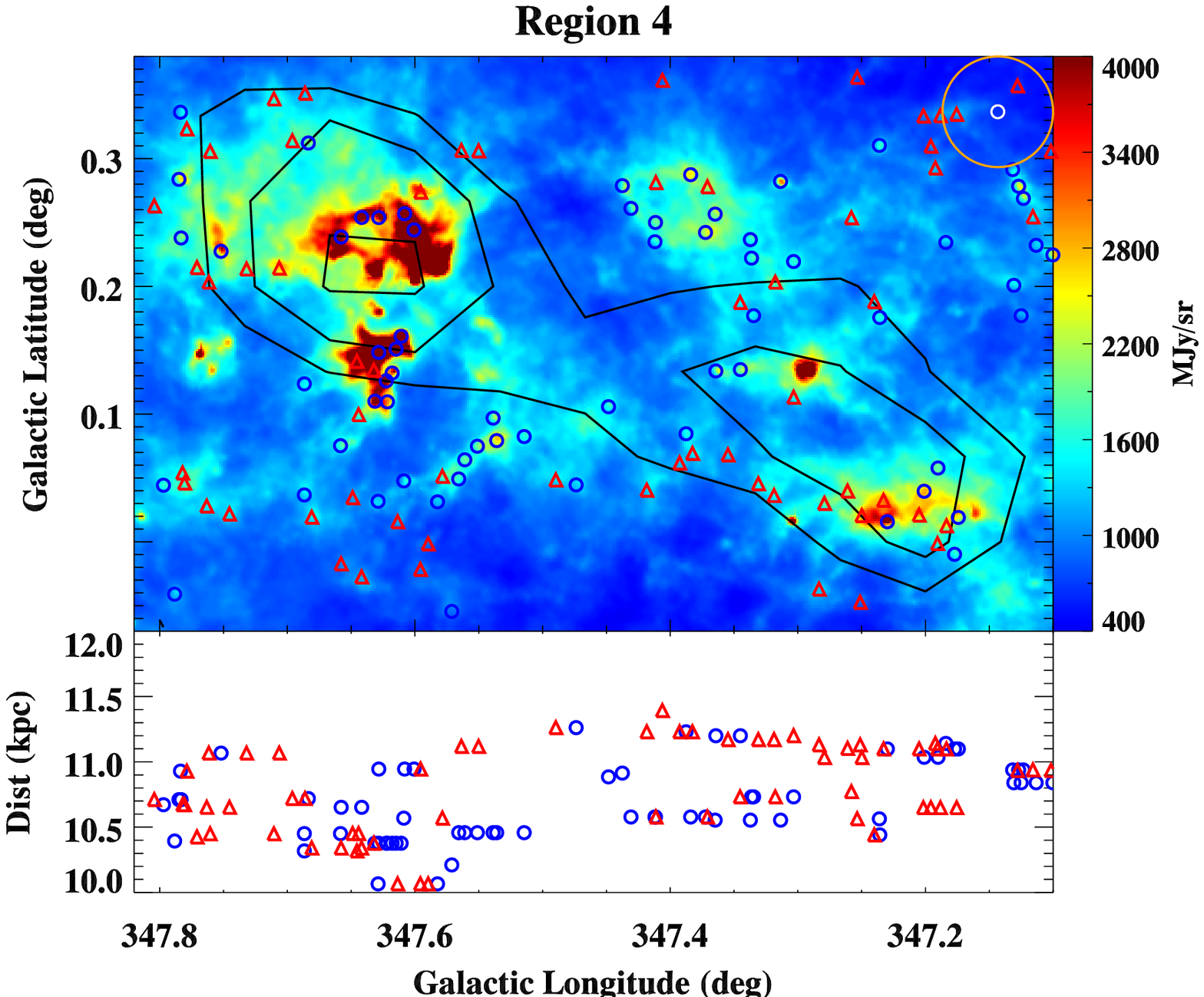}
\includegraphics[width=0.5\textwidth,angle=0]{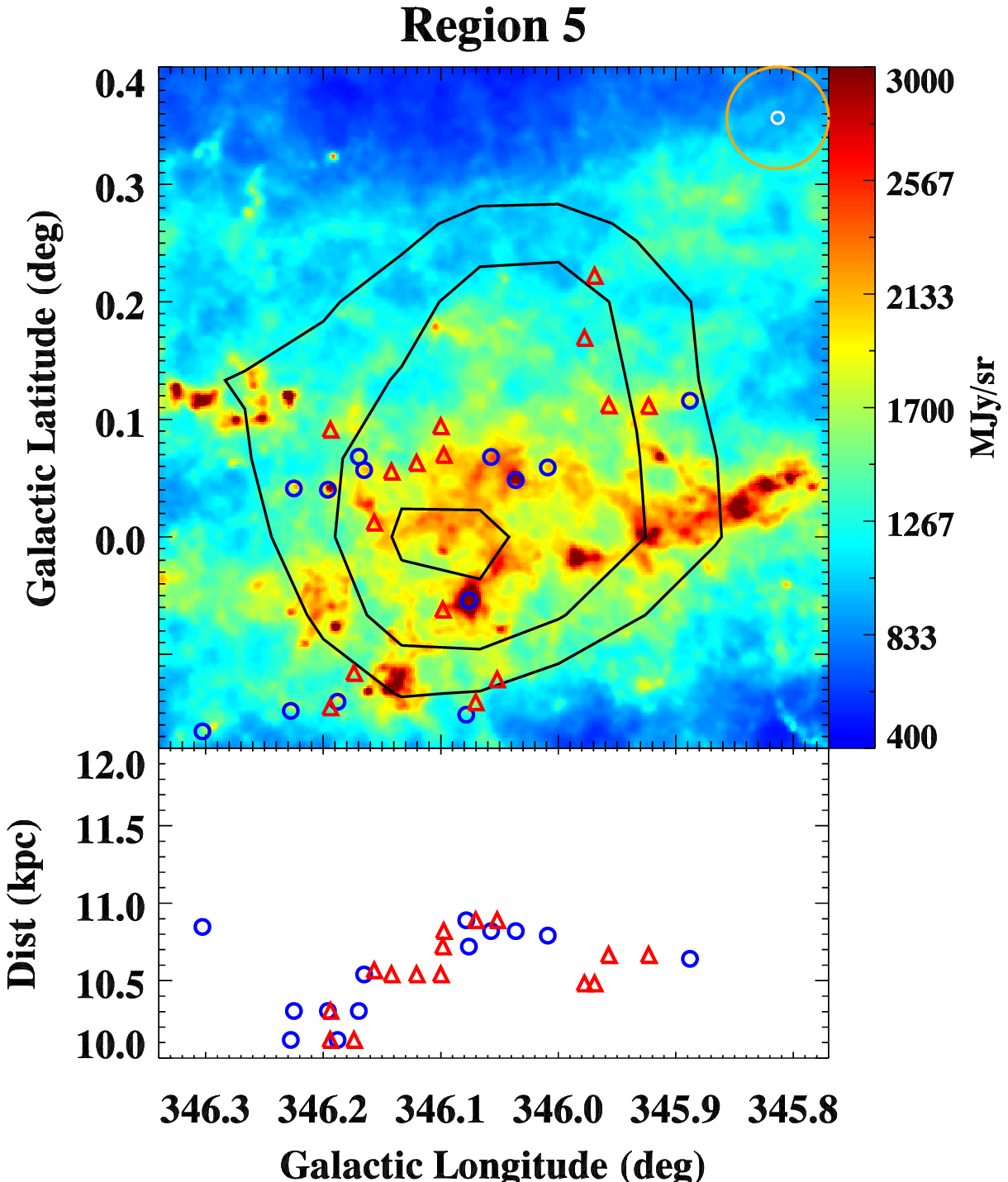}
\caption{SPIRE 250 $\mu$m image of the five giant molecular clouds identified through NANTEN CO contours at the edge of the long Galactic bar in the fourth quadrant (top) and their heliocentric distance as a function of the Galactic longitude (bottom). The black lines show the CO contours at 3$\sigma$, 4$\sigma$ and 5$\sigma$, with $\sigma=3.2$ K. Prestellar clumps belonging to the cloud are plotted in red triangles, while the protostellar objects are shown in blue circles. In the top right hand side the SPIRE beam (white circle) and the NANTEN beam (orange circle) are also shown.  }
\label{fig:sfreg_1}
\end{center}
\end{figure*}

Keeping all those caveats in mind, we proceed to identify the larger star forming complexes in the fourth quadrant and estimate their physical parameters, as other studies have done in the first quadrant. Identification of the star forming regions in the fourth quadrant {employs} ancillary $^{12}$CO(1-0) data from NANTEN. 
We integrated the data cube along the velocities associated to the distance of the bar, which we consider to be between 10 and 12 kpc, that is -60 $<v_{LSR}<$-40 km/s~\citep{Benjamin08}. Then, we drew contours where the signal is higher than 3 RMS of the integrated CO map. 
{We identified with these data five molecular complexes}, named R1 to R5. In order to compare the molecular gas emission with the dust emission, we show 
in Figure~\ref{fig:sfreg_1} the SPIRE 250$\mu$m map of the regions where we overlap CO contours and sources located at the distance of the bar (protostellar  and prestellar objects are both indicated).
We have to consider that the CO integrated map has a lower resolution than the Hi-GAL clumps. Therefore, the CO can trace the gas emission on large scales but not its configuration on smaller sizes. Keeping this in mind, there is a good correspondence between the dust with sources embedded and the molecular gas emissions of the natal cloud.

The physical parameters of each region have been estimated through the methods described in Sections~\ref{sec:counts} and are reported in Table~\ref{tab:sfr_reg_4}, together with the properties of two well studied and very active star forming regions in the first quadrant, W43 and G29.96-0.02, for comparison. 
The SFR of W43 has been thoroughly studied in~\cite{Nguyen11} using the Spitzer IRAC 8$\mu$m database combined with $^{12}$CO and $^{13}$CO observations, though in their case it includes both W43 and G29.96-0.02. They find a value ranging from 0.01~M$_\odot$ yr$^{-1}$ to 0.1~M$_\odot$ yr$^{-1}$ depending on the tracer, with an average efficiency in the range ($1\%-3\%$). While their efficiency values are consistent {with} ours, the SFR estimated with our FIR protostar counting method combined with the~\cite{McKee03} evolutionary model, is lower by around two orders of magnitude. {This might indicate that W43 is populated by a set of already evolved protostars, emitting in the NIR.} When comparing the clouds in the fourth quadrant with W43 and G29.96-0.02, we can see that CFE, $\epsilon_\star,$ and $\Sigma_{SFR}$ seem to be consistent on the two sides of the bar. R2 and R4, being the most extended, show the highest SFR value. The highest $\Sigma_{SFR}$ is also presented by R4, along with a significant $\epsilon_\star$ showing a particularly active area. As a comparison {with R4}, W43 has a higher $\epsilon_\star$ but a lower $\Sigma_{SFR}$. The most efficient region for converting dust into stars is G29.96-0.02 with a $\epsilon_\star$ value two or even three times larger than the clouds in the further  side of the bar. 
Again, the clumps at the bar show efficiencies consistent with the rest of the field, suggesting that the higher star-formation activity in those regions is due to a larger concentration of molecular material {  channeled} from the gravitational potential lines circling around the long Galactic bar, rather than to some triggering event.  

\begin{table*}[!htb]
\begin{center}
\space
\caption{GMC in the 4$^{th}$ quadrant}
\label{tab:sfr_reg_4}
\begin{tabular}{c c c c c c c c c c c c}
\hline
Reg & l & b & $\delta$ & $\alpha$ & d &M$_{\mathrm gas}$ & SFR  & $\Sigma_{\mathrm SFR}$& $\tau_{depl}$ &CFE & $\epsilon_\star$\\
       &(deg) & (deg) &  (pc)  &  & (kpc) & ($10^5$ M$_\odot$) & ($10^{-4}$ M$_\odot$ yr$^{-1}$) & ($10^{-2}$ M$_\odot$ yr$^{-1}$ kpc$^{-2}$) & ($10^{2}$ Myr ) &  & ($\%$) \\
\hline
\hline
R1 &  349.7 & 0.13 & 20.1&  0.40 &11.2 & $0.9$ &  $1.3\pm0.3$  & $9.7\pm 1.9$   & $ {6.9   \pm 1.6}$ & 0.28 & 1.5        \\
R2 & 348.7       & 0.15 & 68.6 & 0.32 & 11.0 & $11.4$ & $9.9\pm 1.9$       & $6.7\pm 1.3$ & $   {11.5  \pm    2} $& 0.19 & 0.7\\
R3 & 347.85 & -0.3 & 29.3 & 1.50 & 10.3 &$2.3$ & $1.7 \pm 0.3$ &     $6.2\pm 1.2$      & $ {13.5  \pm 2 } $ & 0.15 & 0.9\\
R4 & 347.5 & 0.1 & 41.9 & 0.56 & 10.7 &$2.6$ & $6.3\pm1.3$     &     $11.4\pm2.4$     & $ {4.1    \pm 0.8} $ & 0.39 & 1.1\\
R5 & 346.15 & 0.05 & 36.9 & 1.00 & 10.6 &$2.2$ & $1.3\pm0.3$  &    $3.1\pm0.6$      & $ {16.9  \pm  5}  $ & 0.15 & 1.0\\
\hline
W43 & 30.75 & -0.15 & 49.2 & 0.54 & 5.9 & 12.3 &$4.4\pm1.0$ & $5.8\pm 1.3 $& ${27.9    \pm  7} $ &0.08  & 1.5\\
G29.96-0.02 & 29.85 & -0.15 & 31.0 & 1.40 & 6.0 & 1.3 & $1.8\pm0.4$ & $6.1\pm1.2 $ & $ {7.2  \pm 2}$ & 0.24  & 2.8\\
\hline
\end{tabular}
\end{center}
\footnotesize{The first column reports the identifier of the regions, the second and third columns the Galactic coordinates of the approximate centers. All the regions have been approximated as {ellipses}, where the circularized radius ($\delta$) is reported in Column 4 and the {aspect ratio} $\alpha$ (the ratio between the two semi-axis), is shown in Column 5. Column 6 reports the distance, while Column 7 is the $^{12}$CO(1-0) mass inside the larger contour, estimated as described in Section~\ref{sec:gas_mass} while the evolutionary properties are reported in Columns 8 to 12. {  Here we have reported the  SFR of the entire cloud $\Sigma_{\mathrm SFR}$, an estimate of the depletion timescale $\tau_{depl}$ }, the CFE from Equation~\ref{eq:cfe} and $\epsilon_\star$ from Equation~\ref{eq:sfe}. For the regions in the first quadrant the $^{13}$CO(1-0) data from the GRS survey are employed. {  The depletion timescale $\tau_{depl}$ has been computed as the time needed to convert the {entire} cloud mass into stars assuming a constant star-formation rate equal to the estimated SFR.}
}
\end{table*}
\space

\section{Conclusions}\label{sec:conclusions}

We study the physical and evolutionary properties of young clumps at the {tips} of the long Galactic bar by analyzing Herschel/Hi-GAL observations combined with ancillary $^{12}$CO(1-0) NANTEN and $^{13}$CO(1-0) GRS data. The study focuses on two large 2$^\circ$ wide stripes {centered in the GP at $b=0^\circ$} and covering the latitude ranges $19^\circ<\ell<33^\circ$ and $340^\circ<\ell<350^\circ$. The observed fields cross the ends of the bar and the beginning of the Galactic spiral arms. According to our study, the source distribution close to the Galactic long bar indicates that the tips are located at $24^\circ<\ell<31^\circ$ and $346^\circ<\ell<350^\circ$ (see section~\ref{sectionclass}). In this paper, an independent evaluation of Hi-GAL source distances was carried out with the BGPS survey as reference. The distance correspondences are as good as 70$\%$ on sources where a positional match is found (see section~\ref{sec:distance}). Furthermore, our study shows that the physical properties, such as temperatures, masses and sizes, of prestellar and protostellar clumps located at the {tips} of the bar are consistent with {sources in the same evolutionary stage located in background and foreground regions} (see section~\ref{sectionclass}).

We estimate the SFR and $\Sigma_{\mathrm SFR}$ by adapting a star-counting method, based on the~\cite{McKee03} model, to an evolutionary framework where each clump collapses into a star cluster, consistently with observations of the massive star-formation process. The analysis {is carried out at the tips of the Galactic bar and in background and foreground regions}. {The main properties are} reported in this paper as a function of the distance from the GC and based on their location with respect to the tips of the bar. Our study shows that most of the star-formation activity in the first quadrant occurs in the heliocentric distance range 5 - 11 kpc where the line of sight crosses first the tip of the bar and then the Perseus arm. The star-formation rate density at the tip is $1.2\pm0.3\;10^{-3}\mathrm{\frac{M_\odot}{yr\;kpc^2}}$, while the star-formation rate density on the entire field of view in the first quadrant is $0.9\pm0.2\;10^{-3}\mathrm{\frac{M_\odot}{yr\;kpc^2}}$. These values are confirmed by the star-formation rate calculation method based on infrared luminosities, which provides consistent results (section~\ref{sec:sfr}). Most of the star-formation activity in the fourth quadrant occurs at $\sim2-3$ kpc, where local star forming regions are situated, and at $\sim$11 kpc, where the further tip of the bar is located. The star-formation rate density at the tip is $1.5\pm0.3\;10^{-3}\mathrm{\frac{M_\odot}{yr\;kpc^2}}$, while the star-formation rate density on the entire field of view in the fourth quadrant is $0.8\pm0.2\;10^{-3}\mathrm{\frac{M_\odot}{yr\;kpc^2}}$. As for the first quadrant, this results are confirmed by the method based on infrared luminosities (section~\ref{sec:sfr}).

Our analysis shows also that the conversion efficiencies, which indicate the percentage amount of material converted into stars, is approximately 0.8$\%$ in the first quadrant and 0.5$\%$ in the fourth quadrant and does not show a significant difference in proximity of the bar (section~\ref{sec:sfe}).

A detailed study {on the evolutionary properties of five giant molecular complexes was also conducted. Those complexes were identified through CO contours at the tip in the fourth Galactic quadrant}. According to our analysis, the complexes show star-formation rate densities in the range $3.1-11.4\;10^{-2}\mathrm{\frac{M_\odot}{yr\;kpc^2}}$. Despite the high star-formation rate densities, the conversion efficiencies are in the range 0.7-1.5$\%$ which is fully consistent with the values in the surrounding regions (section~\ref{sec:q4}).

{Our analysis shows that the star-formation activity at the tips of the bar is enhanced with respect to background and foreground regions, with active giant molecular complexes both in the first and fourth quadrant. However, the conversion efficiency does not show significant changes at the bar or in the molecular complexes with respect to the other fields analyzed. This result suggests that the enhanced star-formation activity at the bar occurs mostly because of a higher concentration of dust and molecular material, which are carried in the area due to the particular configuration of the gravitational potential, rather than to a specific triggering process. }

\section*{Acnowledgments}
The authors wish to thank the referee, Prof. Neal Evans, for his careful reading of the manuscript and his useful comments that have helped us to improve further the content 
of the paper. MV thanks Science [$\&$] Technology corp. for their support with data processing. The research activity of DE and ES has been supported by the VIALACTEA Project, a Collaborative Project under Framework Programme 7 of the European Union funded under Contract $\#$607380.

\bibliographystyle{aa}

\newpage

\end{document}